\documentclass[%
 reprint, 
 superscriptaddress,
 amsmath,amssymb,
 longbibliography,
  pra,
 aps,
]{revtex4-2}

\usepackage{graphicx}   
\usepackage{dcolumn}    
\usepackage{bm}         
\usepackage{hyperref}   
\usepackage{graphicx}
\usepackage[caption=false]{subfig}
\usepackage{tikz}
\usepackage{float}
\usepackage{placeins}
\usepackage{adjustbox} 
\usepackage{multirow}
\usepackage[normalem]{ulem} 
\usetikzlibrary{
    shapes.geometric, 
    arrows.meta, 
    positioning, 
    fit, 
    backgrounds,
    }
\usepackage{comment}
\hypersetup{
colorlinks=true,
linkcolor=blue,
filecolor=blue,
citecolor=blue,  
urlcolor=blue,
}

\newcommand{\zw}[1]{ { \color{red} \footnotesize (\textsf{zw}) \textsf{\textsl{#1}} }}

\newcommand{\revision}[1]{ {\color{blue} \footnotesize (\textsf{revision}) \textsf{\textsl{#1}} }}

\newcommand{\rev}[1]{{\color{black}#1}}

\newcommand{\revisiontwo}[1]{ {\color{green} \footnotesize (\textsf{revision}) \textsf{\textsl{#1}} }}

\newcommand{\qldpc}[5]{%
  $[\![#1,#2,#3]\!]_{\!(#4,#5)}$%
}

\usepackage{xcolor,colortbl}

\definecolor{Gray}{gray}{0.85}
\definecolor{LightCyan}{rgb}{0.88,1,1}
\newcolumntype{a}{>{\columncolor{Gray}}l}

\begin{document}

\title{Discovering highly efficient low-weight quantum error-correcting codes\\ with reinforcement learning}

\author{Austin Yubo He}
\email{austinhe@utexas.edu}
\affiliation{
The University of Texas at Austin, Austin, TX 78712, USA
}
\affiliation{
University of California, Berkeley, Berkeley, CA 94720, USA
}
\author{Zi-Wen Liu}
\email{zwliu0@tsinghua.edu.cn}
\affiliation{Yau Mathematical Sciences Center, Tsinghua University, Beijing, 100084, China }

\date{\today}

\begin{abstract}

\rev{The realization of scalable fault-tolerant quantum computing is expected to hinge on quantum error-correcting codes. 
In the quest for more efficient quantum fault tolerance, a critical code parameter is the weight of measurements that extract information about errors to enable error correction: as higher measurement weights require higher implementation costs and introduce more errors, it is important in code design to optimize measurement weight.}  
 \rev{This underlies the surging interest in quantum low-density parity-check (qLDPC) codes, the study of which has primarily focused on the asymptotic (large-code-limit) properties.} 
  \rev{In this work, we introduce a versatile and computationally efficient approach to stabilizer code weight reduction based on reinforcement learning (RL), which produces new low-weight codes that substantially outperform the state of the art  in practically relevant parameter regimes, extending significantly beyond previously accessible small distances.} 
 \rev{For example,  {our approach demonstrates savings in physical qubit overhead  compared to existing results by 1 to 2 orders of magnitude for weight 6 codes} and brings the overhead into a feasible range for near-future experiments.}  
 \rev{We also investigate the interplay between code parameters using our RL framework, offering new insights into the potential efficiency and power of practically viable coding strategies.} \rev{Overall, our results demonstrate how RL can effectively advance the crucial yet challenging problem of quantum code discovery and thereby facilitating a faster path to the practical implementation of fault-tolerant quantum technologies.}  
\end{abstract}

\maketitle


\section{Introduction}
\label{sec:intro}

\rev{Quantum information processing offers promising potential for revolutionary advantages over conventional methods in computation and various other types of technologies~\cite{Nielsen_Chuang_2010,Shor_1997,Lloyd:1996aai,Giovannetti:2011chh}}.  However, \rev{a fundamental obstacle stands in the way: quantum systems and their manipulation are inherently prone to a wide variety of noise and errors, necessitating efficient fault tolerance strategies that maintain the protection of quantum information when all components may be faulty, in order to make quantum advantages practically scalable and unlock their full potential. Quantum error-correcting (QEC) codes provide a pathway to efficient fault tolerance, positioning them as a pivotal field of research in quantum information~\cite{shor1995qec,gottesman1997stabilizer,Shor1996FaultTolerant,Campbell_2017}. Furthermore, they have recently made a profound impact on physics~\cite{KITAEV20032,zeng2018quantuminformationmeetsquantum,Almheiri2015,YYGL2024}, underscoring their fundamental importance.} 

\rev{Stabilizer codes serve as a canonical framework for QEC codes, enabling logical qubits to be encoded in more physical qubits in a way that information about errors needed for correction can be inferred through measuring certain parity-check (stabilizer) operators}~\cite{gottesman1997stabilizer}. 
\rev{Since measurements of higher weight (the size of nontrivial support) typically require larger circuits and qubit overhead, which can make experimental executions significantly more difficult and introduce more errors, it is crucial to minimize the check weight  in code design for practical quantum computing.}
\rev{Note that lower check weight is also a key motivation behind the interest in various generalized coding schemes such as subsystem codes~\cite{Poulin_2005,Aliferis07:subsystem} and dynamical codes~\cite{Hastings2021dynamically,fu2024errorcorrectiondynamicalcodes}.} 

\rev{In line with this, constraining the check weight (as well as the degree) to be asymptotically $O(1)$---that is, bounded by a constant as the code length grows---gives rise to the so-called quantum low-density parity-check (qLDPC) codes which have attracted intensive interest as a promising scheme for fault tolerance with low overhead~\cite{10.5555/2685179.2685184, breuckmann_qldpc_2021, cohen_low-overhead_2022, iolius_almost-linear_2024}.} 
\rev{In particular, general qLDPC codes can achieve substantially better code parameters and fault tolerance efficiency~\cite{breuckmann_qldpc_2021,Bravyi_2024,xu_constant-overhead_2024} than those with geometric connectivity constraints including the surface code, which has long been regarded the leading scheme for implementing fault tolerance~\cite{KITAEV20032,Raussendorf_surface,fowler2012surface,acharya_suppressing_2023,acharya_quantum_2024}. The need for substantial long-range connectivity to overcome geometric barriers for code parameters~\cite{BravyiTerhal_2009,BPT,baspin_quantifying_2022, baspin_connectivity_2022, baspin_improved_2023} poses a significant obstacle to capitalizing on the advantages of qLDPC codes. However, recent remarkable advances in quantum computing with the reconfigurable atom array platform~\cite{Bluvstein_2023} bring hope for alleviating this difficulty, potentially establishing qLDPC code-based fault tolerance schemes  as mainstream. Driven largely by theoretical interest, intensive study has been devoted to the constructions of qLDPC code families with desirable asymptotic parameters in the infinite code length limit (see e.g.~Ref.~\cite{breuckmann_qldpc_2021} for a slightly outdated review). Notably, there has been a recent surge of breakthroughs in achieving asymptotically `good' qLDPC code families that simultaneously attain optimal scalings of both code rate and distance~\cite{panteleev2022asymptoticallygoodquantumlocally,Leverrier_2022,dinur2022goodquantumldpccodes}. However, in practically relevant finite-size regimes, these asymptotic code constructions are not expected to exhibit good parameters and usually feature stabilizers with relatively high (though finite) weights too demanding for actual implementation. The optimization of finite-scale code design requires specialized approaches yet has received little attention, despite its evident importance for the practical development of quantum hardware. In particular, recent experimental progress~\cite{Bluvstein_2023,acharya_quantum_2024} suggests that QEC codes with a distance of several tens are crucial in the coming years for the development of fault-tolerant hardware, whereas existing code design approaches (ranging from e.g.~greedy algorithms, constraint satisfaction, exhaustive search, evolutionary algorithms, to reinforcement learning~\cite{tremblay_finite-rate_2023, olle_simultaneous_2024, Mauron_2024, su_discovery_2023, nautrup_optimizing_2019, zeng_approximate_2023, tandeitnik_evolving_2024, webster_engineering_2024}) generally struggle to exceed single-digit distances.}



\rev{In the pursuit of optimizing QEC codes, an effective strategy is known as weight reduction, which involves algorithms that aim to decrease check weight while maintaining other code properties like rate and distance, at the cost of  physical qubit overhead.} 
\rev{Hastings first proposed a weight reduction method for CSS codes~\cite{hastings_weight_2016,hastings_quantum_2023} primarily focusing on the asymptotic setting, reducing the check weight and degree to $O(1)$.} 
Subsequent work by Sabo et al.~\cite{sabo_weight-reduced_2024} extended the idea to finite-size regimes for product codes, demonstrating a modified weight reduction method that can be applied with significantly lower qubit overhead on relatively small codes and better performance when implemented on a cluster state architecture using GKP qubits~\cite{tzitrin_fault-tolerant_2021}. \rev{Nevertheless, as we shall demonstrate, their method is still far from optimal, typically entailing qubit overhead that can be significantly improved especially larger-size regimes that are crucial for future applications.}

\rev{In this work, we present a remarkably effective and general scheme for discovering low-weight QEC codes based on a novel reinforcement learning (RL)~\cite{10.5555/980651.980663} scheme for weight reduction. This addresses the previously recognized difficulty of learning relatively large qLDPC codes as explained in, e.g.,~Ref.~\cite{olle_simultaneous_2024}.}
\rev{Our results reveal an essential insight that decreasing weight with distance constraints is a significantly more approachable problem compared to increasing distance with weight constraints, especially for learning methods.} 
\rev{Specifically, a major obstacle that restricts the scale of previous code design approaches
is that distance is a high-complexity property depending on the hamming weight of all non-trivial logical operators, and must be computed through exhaustive search or sampling \cite{kapshikar_hardness_2023}, while weight can be directly calculated from the parity check matrix.} 
\rev{Note that several works~\cite{nautrup_optimizing_2019,Sweke_2020,su_discovery_2023,  zeng_approximate_2023,olle_simultaneous_2024, Mauron_2024, sivak2024optimizationdecoderpriorsaccurate,freire2025optimizinghypergraphproductcodes} have explored the use of RL in QEC code design from various other perspectives including distance, threshold, logical error rate, decoding, and error adaptation, while this work highlights the prominence of check weight.}


\rev{We specifically demonstrate the effectiveness of our approach using hypergraph product codes as base codes.
Notably, we find that our RL model consistently achieves significantly smaller physical qubit overhead---up to 73x in the best case---compared to previous weight reduction methods. Moreover, it learns to design codes with distances up to 4x more than previous approaches applying RL methods to code design. Altogether, our RL-based scheme enables the discovery of numerous new low-weight codes with high distances of up to $d\approx 35$ that requires significantly less physical qubits than existing constructions. 
This brings the efficiency of codes into a feasible range for near-term quantum devices with up to a few thousand physical qubits, and represents the first demonstration of RL-designed codes with parameters sufficiently good to be practically useful. 
}


\begin{figure}

\tikzset{every picture/.style={line width=0.75pt}} 

\begin{tikzpicture}[x=0.75pt,y=0.75pt,yscale=-1,xscale=1]

\draw   (85.99,115.57) .. controls (80.84,115.63) and (76.62,111.33) .. (76.58,105.97) .. controls (76.53,100.62) and (80.67,96.23) .. (85.82,96.18) .. controls (90.98,96.13) and (95.19,100.43) .. (95.24,105.78) .. controls (95.28,111.14) and (91.14,115.52) .. (85.99,115.57) -- cycle ;
\draw   (112.11,115.31) .. controls (106.95,115.36) and (102.74,111.06) .. (102.69,105.71) .. controls (102.65,100.35) and (106.79,95.97) .. (111.94,95.92) .. controls (117.09,95.86) and (121.31,100.16) .. (121.35,105.52) .. controls (121.4,110.87) and (117.26,115.26) .. (112.11,115.31) -- cycle ;
\draw   (33.76,116.1) .. controls (28.61,116.16) and (24.39,111.86) .. (24.35,106.5) .. controls (24.3,101.15) and (28.44,96.76) .. (33.59,96.71) .. controls (38.74,96.66) and (42.96,100.95) .. (43,106.31) .. controls (43.05,111.67) and (38.91,116.05) .. (33.76,116.1) -- cycle ;
\draw   (59.87,115.84) .. controls (54.72,115.89) and (50.51,111.59) .. (50.46,106.24) .. controls (50.42,100.88) and (54.56,96.5) .. (59.71,96.44) .. controls (64.86,96.39) and (69.07,100.69) .. (69.12,106.05) .. controls (69.16,111.4) and (65.03,115.79) .. (59.87,115.84) -- cycle ;
\draw    (130,70) -- (218,70) ;
\draw [shift={(220,70)}, rotate = 180] [color={rgb, 255:red, 0; green, 0; blue, 0 }  ][line width=0.75]    (10.93,-3.29) .. controls (6.95,-1.4) and (3.31,-0.3) .. (0,0) .. controls (3.31,0.3) and (6.95,1.4) .. (10.93,3.29)   ;
\draw   (230,136.15) -- (230.01,119.55) -- (246.28,119.56) -- (246.26,136.17) -- cycle ;
\draw   (230.03,102.93) -- (230.05,86.33) -- (246.31,86.34) -- (246.3,102.95) -- cycle ;
\draw   (230.07,69.72) -- (230.08,53.11) -- (246.35,53.12) -- (246.33,69.73) -- cycle ;
\draw    (246.3,94.65) -- (287.18,22.66) ;
\draw    (246.3,94.65) -- (287.03,76.23) ;
\draw    (246.27,127.87) -- (286.97,160.31) ;
\draw    (246.27,127.87) -- (287.04,133.42) ;
\draw [color={rgb, 255:red, 0; green, 0; blue, 0 }  ,draw opacity=1 ]   (246.27,127.87) -- (286.94,105.48) ;
\draw [color={rgb, 255:red, 117; green, 212; blue, 9 }  ,draw opacity=1 ]   (246.27,127.87) -- (287.03,76.23) ;
\draw    (246.3,94.65) -- (287.04,133.42) ;
\draw    (246.3,94.65) -- (286.97,160.31) ;
\draw    (246.34,61.43) -- (287.11,49.56) ;
\draw    (246.34,61.43) -- (287.03,76.23) ;
\draw    (246.34,61.43) -- (286.94,105.48) ;
\draw [color={rgb, 255:red, 208; green, 2; blue, 27 }  ,draw opacity=1 ]   (246.34,61.43) -- (286.97,160.31) ;
\draw   (73.02,170) .. controls (67.87,170.05) and (63.65,165.75) .. (63.61,160.4) .. controls (63.56,155.04) and (67.7,150.66) .. (72.85,150.6) .. controls (78,150.55) and (82.22,154.85) .. (82.26,160.21) .. controls (82.31,165.56) and (78.17,169.95) .. (73.02,170) -- cycle ;
\draw   (44.65,169.36) .. controls (39.5,169.42) and (35.29,165.12) .. (35.24,159.76) .. controls (35.2,154.41) and (39.34,150.02) .. (44.49,149.97) .. controls (49.64,149.92) and (53.85,154.22) .. (53.9,159.57) .. controls (53.95,164.93) and (49.81,169.31) .. (44.65,169.36) -- cycle ;
\draw   (101.37,169.71) .. controls (96.22,169.76) and (92.01,165.47) .. (91.96,160.11) .. controls (91.92,154.75) and (96.05,150.37) .. (101.21,150.32) .. controls (106.36,150.26) and (110.57,154.56) .. (110.62,159.92) .. controls (110.66,165.27) and (106.52,169.66) .. (101.37,169.71) -- cycle ;
\draw   (85.82,64.93) .. controls (80.67,64.99) and (76.53,69.41) .. (76.58,74.81) .. controls (76.62,80.22) and (80.84,84.55) .. (85.99,84.5) .. controls (91.14,84.45) and (95.28,80.02) .. (95.24,74.62) .. controls (95.19,69.22) and (90.97,64.88) .. (85.82,64.93) -- cycle ;
\draw   (111.94,64.67) .. controls (106.79,64.72) and (102.65,69.14) .. (102.69,74.55) .. controls (102.74,79.95) and (106.95,84.29) .. (112.11,84.24) .. controls (117.26,84.18) and (121.4,79.76) .. (121.35,74.36) .. controls (121.3,68.95) and (117.09,64.62) .. (111.94,64.67) -- cycle ;
\draw   (33.59,65.46) .. controls (28.44,65.52) and (24.3,69.94) .. (24.35,75.34) .. controls (24.39,80.75) and (28.61,85.08) .. (33.76,85.03) .. controls (38.91,84.98) and (43.05,80.55) .. (43,75.15) .. controls (42.96,69.75) and (38.74,65.41) .. (33.59,65.46) -- cycle ;
\draw   (59.71,65.2) .. controls (54.56,65.25) and (50.42,69.67) .. (50.46,75.08) .. controls (50.51,80.48) and (54.72,84.82) .. (59.87,84.77) .. controls (65.03,84.71) and (69.17,80.29) .. (69.12,74.89) .. controls (69.07,69.48) and (64.86,65.14) .. (59.71,65.2) -- cycle ;
\draw   (71.93,10.3) .. controls (66.78,10.35) and (62.64,14.77) .. (62.68,20.17) .. controls (62.73,25.58) and (66.94,29.92) .. (72.1,29.86) .. controls (77.25,29.81) and (81.39,25.39) .. (81.34,19.98) .. controls (81.29,14.58) and (77.08,10.24) .. (71.93,10.3) -- cycle ;
\draw   (43.58,11.51) .. controls (38.43,11.57) and (34.29,15.99) .. (34.34,21.39) .. controls (34.38,26.8) and (38.6,31.13) .. (43.75,31.08) .. controls (48.9,31.03) and (53.04,26.6) .. (52.99,21.2) .. controls (52.95,15.8) and (48.73,11.46) .. (43.58,11.51) -- cycle ;
\draw   (100.28,10.01) .. controls (95.13,10.06) and (90.99,14.48) .. (91.04,19.89) .. controls (91.09,25.29) and (95.3,29.63) .. (100.45,29.58) .. controls (105.6,29.52) and (109.74,25.1) .. (109.7,19.7) .. controls (109.65,14.29) and (105.44,9.95) .. (100.28,10.01) -- cycle ;
\draw    (43.75,31.08) -- (59.71,65.2) ;
\draw    (72.1,29.86) -- (85.82,64.93) ;
\draw    (100.45,29.58) -- (111.94,64.67) ;
\draw    (72.1,29.86) -- (59.71,65.2) ;
\draw    (100.45,29.58) -- (85.82,64.93) ;
\draw    (43.75,31.08) -- (33.59,65.46) ;
\draw    (101.21,150.32) -- (85.99,115.58) ;
\draw    (72.85,150.6) -- (59.87,115.84) ;
\draw    (44.49,149.97) -- (33.76,116.1) ;
\draw    (72.85,150.6) -- (85.99,115.58) ;
\draw    (44.49,149.97) -- (59.87,115.84) ;
\draw    (101.21,150.32) -- (112.1,115.31) ;
\draw    (44.49,149.97) -- (85.99,115.58) ;
\draw    (44.49,149.97) -- (112.1,115.31) ;
\draw    (33.59,65.46) -- (100.45,29.58) ;
\draw    (33.59,65.46) -- (72.1,29.86) ;
\draw    (59.71,65.2) -- (100.45,29.58) ;
\draw    (85.82,64.93) -- (43.75,31.08) ;
\draw    (111.94,64.67) -- (72.1,29.86) ;
\draw    (111.94,64.67) -- (43.75,31.08) ;
\draw    (72.85,150.6) -- (33.76,116.1) ;
\draw    (112.1,115.31) -- (72.85,150.6) ;
\draw    (59.87,115.84) -- (101.21,150.32) ;
\draw    (33.76,116.1) -- (101.21,150.32) ;
\draw    (220,130) -- (132,130) ;
\draw [shift={(130,130)}, rotate = 360] [color={rgb, 255:red, 0; green, 0; blue, 0 }  ][line width=0.75]    (10.93,-3.29) .. controls (6.95,-1.4) and (3.31,-0.3) .. (0,0) .. controls (3.31,0.3) and (6.95,1.4) .. (10.93,3.29)   ;
\draw   (296.42,12.78) .. controls (291.27,12.83) and (287.13,17.25) .. (287.18,22.66) .. controls (287.23,28.06) and (291.44,32.4) .. (296.59,32.35) .. controls (301.74,32.29) and (305.88,27.87) .. (305.84,22.47) .. controls (305.79,17.06) and (301.58,12.72) .. (296.42,12.78) -- cycle ;
\draw   (296.35,39.69) .. controls (291.2,39.74) and (287.06,44.16) .. (287.11,49.56) .. controls (287.16,54.97) and (291.37,59.31) .. (296.52,59.25) .. controls (301.67,59.2) and (305.81,54.78) .. (305.77,49.37) .. controls (305.72,43.97) and (301.51,39.63) .. (296.35,39.69) -- cycle ;
\draw   (296.27,66.35) .. controls (291.12,66.4) and (286.98,70.83) .. (287.03,76.23) .. controls (287.07,81.63) and (291.29,85.97) .. (296.44,85.92) .. controls (301.59,85.86) and (305.73,81.44) .. (305.68,76.04) .. controls (305.64,70.63) and (301.42,66.3) .. (296.27,66.35) -- cycle ;
\draw   (296.19,95.6) .. controls (291.03,95.65) and (286.9,100.08) .. (286.94,105.48) .. controls (286.99,110.88) and (291.2,115.22) .. (296.35,115.17) .. controls (301.51,115.11) and (305.64,110.69) .. (305.6,105.29) .. controls (305.55,99.88) and (301.34,95.55) .. (296.19,95.6) -- cycle ;
\draw   (296.29,123.54) .. controls (291.13,123.59) and (287,128.02) .. (287.04,133.42) .. controls (287.09,138.82) and (291.3,143.16) .. (296.45,143.11) .. controls (301.61,143.05) and (305.74,138.63) .. (305.7,133.23) .. controls (305.65,127.82) and (301.44,123.49) .. (296.29,123.54) -- cycle ;
\draw   (296.22,150.43) .. controls (291.07,150.48) and (286.93,154.91) .. (286.97,160.31) .. controls (287.02,165.72) and (291.23,170.05) .. (296.39,170) .. controls (301.54,169.95) and (305.68,165.52) .. (305.63,160.12) .. controls (305.58,154.72) and (301.37,150.38) .. (296.22,150.43) -- cycle ;
\draw [color={rgb, 255:red, 224; green, 224; blue, 224 }  ,draw opacity=1 ]   (33.59,96.71) -- (33.76,85.03) ;
\draw [color={rgb, 255:red, 224; green, 224; blue, 224 }  ,draw opacity=1 ]   (59.71,96.44) -- (59.87,84.77) ;
\draw [color={rgb, 255:red, 224; green, 224; blue, 224 }  ,draw opacity=1 ]   (85.82,96.18) -- (85.99,84.5) ;
\draw [color={rgb, 255:red, 224; green, 224; blue, 224 }  ,draw opacity=1 ]   (111.94,95.91) -- (112.11,84.24) ;
\draw [color={rgb, 255:red, 224; green, 224; blue, 224 }  ,draw opacity=1 ]   (111.94,95.91) -- (85.99,84.5) ;
\draw [color={rgb, 255:red, 224; green, 224; blue, 224 }  ,draw opacity=1 ]   (111.94,95.91) -- (59.87,84.77) ;
\draw [color={rgb, 255:red, 224; green, 224; blue, 224 }  ,draw opacity=1 ]   (111.94,95.91) -- (33.76,85.03) ;
\draw [color={rgb, 255:red, 224; green, 224; blue, 224 }  ,draw opacity=1 ]   (85.99,84.5) -- (59.71,96.44) ;
\draw [color={rgb, 255:red, 224; green, 224; blue, 224 }  ,draw opacity=1 ]   (85.99,84.5) -- (33.59,96.71) ;
\draw [color={rgb, 255:red, 224; green, 224; blue, 224 }  ,draw opacity=1 ]   (112.11,84.24) -- (85.82,96.18) ;
\draw [color={rgb, 255:red, 224; green, 224; blue, 224 }  ,draw opacity=1 ]   (112.11,84.24) -- (33.59,96.71) ;
\draw [color={rgb, 255:red, 224; green, 224; blue, 224 }  ,draw opacity=1 ]   (59.87,84.77) -- (33.59,96.71) ;
\draw [color={rgb, 255:red, 224; green, 224; blue, 224 }  ,draw opacity=1 ]   (33.76,85.03) -- (59.71,96.44) ;
\draw [color={rgb, 255:red, 224; green, 224; blue, 224 }  ,draw opacity=1 ]   (33.76,85.03) -- (85.82,96.18) ;
\draw [color={rgb, 255:red, 224; green, 224; blue, 224 }  ,draw opacity=1 ]   (59.71,96.44) -- (112.11,84.24) ;
\draw [color={rgb, 255:red, 224; green, 224; blue, 224 }  ,draw opacity=1 ]   (59.87,84.77) -- (85.82,96.18) ;
\draw    (220,100) -- (132,100) ;
\draw [shift={(130,100)}, rotate = 360] [color={rgb, 255:red, 0; green, 0; blue, 0 }  ][line width=0.75]    (10.93,-3.29) .. controls (6.95,-1.4) and (3.31,-0.3) .. (0,0) .. controls (3.31,0.3) and (6.95,1.4) .. (10.93,3.29)   ;

\draw (328,55) node [anchor=north west][inner sep=0.75pt]  [font=\normalsize,rotate=-90] [align=left] {{\fontfamily{ptm}\selectfont Environment}};
\draw (150,82) node [anchor=north west][inner sep=0.75pt]  [font=\normalsize] [align=left] {{\fontfamily{ptm}\selectfont State: $H$}};
\draw (135,112) node [anchor=north west][inner sep=0.75pt]  [font=\normalsize] [align=left] {{\fontfamily{ptm}\selectfont Reward: $d,w,q$}};
\draw (130,52) node [anchor=north west][inner sep=0.75pt]  [font=\normalsize] [align=left] {{\fontfamily{ptm}\selectfont Action: +/- edge}};

\draw (2,130) node [anchor=north west][inner sep=0.75pt]  [font=\normalsize,rotate=-270] [align=left] {{\fontfamily{ptm}\selectfont Policy: 
$\pi_\theta(a|s)$ }};
\end{tikzpicture}
\caption{\rev{{An illustration of our RL scheme.} The RL agent (left) maintains a \emph{policy network} that, given the \emph{state} of the Tanner graph, selects an \emph{action} of adding or removing an edge. The environment (right) updates the graph accordingly and returns a  \emph{reward} based on the code’s new distance and weight. This reward signal is then used to update the policy network, guiding the agent toward better code designs.}  
} 
\label{fig:rl_diagram}
\end{figure}
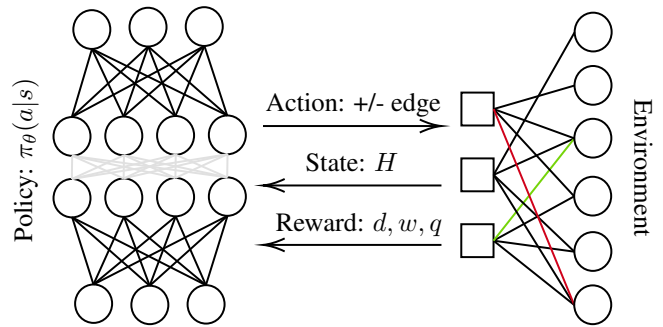


\section{Reinforcement learning framework for weight reduction}
\label{sec:methods}

Reinforcement learning (RL) is a \rev{widely used machine learning paradigm} in which an agent interacts with an environment by selecting actions and receiving feedback in the form of reward signals. The agent iteratively refines its decision-making policy to maximize cumulative rewards \cite{10.5555/980651.980663}. This paradigm has proven successful in a number of problems in physics \cite{sgroi_reinforcement_2021, melnikov_setting_2020, bolens_reinforcement_2021, boccardo_reinforcement_2024, borah_measurement-based_2021}, as well as problems with exceptionally large and complex decision spaces \cite{silver_mastering_2016} and thus naturally lends itself to the combinatorial nature of code design.

As \rev{illustrated} in Fig.~\ref{fig:rl_diagram}, the environment of our weight reduction framework is defined by the Tanner graph of a stabilizer code. The agent can \textit{remove} or \textit{add} edges in the Tanner graph, and its objective is to minimize the maximum degree of variable and check nodes while preserving the code distance. 

We utilize the Proximal Policy Optimization (PPO) algorithm with action masking to efficiently explore the action space~\cite{schulman_proximal_2017}. PPO balances exploration and exploitation by constraining policy updates, ensuring stable learning. The objective function is
\begin{equation}
    \mathcal{L}^{\text{CLIP}}(\theta) = \mathbb{E}_t \left[ \min \left( r_t(\theta) \hat{A}_t, \ \text{clip}\left(r_t(\theta), 1 - \epsilon, 1 + \epsilon\right) \hat{A}_t \right) \right],
\end{equation}
where \( r_t(\theta) = \frac{\pi_\theta(a_t|s_t)}{\pi_{\theta_{\text{old}}}(a_t|s_t)} \) is the probability ratio, \( \hat{A}_t \) is the advantage estimate, and \( \epsilon \) controls the clipping range.

\rev{Then we design the} reward function to guide our Reinforcement Learning (RL) agent in optimizing Tanner graphs for stabilizer codes. The function balances node degree reduction with code distance preservation, ensuring robust error-correcting capabilities. 
We define local reward functions \(R_v(\kappa_v)\) and \(R_c(\kappa_c)\) for variable and check nodes, respectively, by assigning fixed rewards for degrees up to \(V_{\max}\), and an exponential penalty for degrees above \(V_{\max}\):
\begin{equation}
    R_v(\kappa_v) 
    \;=\;
    \begin{cases}
        C_1, & \kappa_v = 1,\\
        C_2, & \kappa_v = 2,\\
        \vdots & \\
        C_{V_{\max}}, & \kappa_v = V_{\max},\\
        \exp\!\bigl[-\lambda\,(\kappa_v - V_{\max})\bigr], & \kappa_v > V_{\max},
    \end{cases}
    \label{eq:Rv-kappa}
\end{equation}
where \(\lambda\) controls the severity of the penalty for large degrees, and the constants \(C_1, C_2, \ldots, C_{V_{\max}}\) reflect the relative desirability of each integer degree within the allowable range. A similar function \(R_c(\kappa_c)\) can be defined for check nodes. In our demonstration using \rev{hypergraph product} codes we have $V_{\max}=3$. The gap between the values of $C_3$ and $C_2$ was set to be close, with both around 0.7 to 1.0, while $C_1$ was set to be a smaller value around 0.1 to 0.5. The precise values were adjusted empirically based on training results to encourage degree distributions to avoid under-utilizing check operators.

We min-max normalize each local reward $R_v(\kappa_v), R_c(\kappa_c)$ and the distance-based terms $d, \Delta d$ to fall in the interval $[0,1]$. \rev{Specifically, for any quantity $X$ that lies in $[X_{\min},X_{\max}]$, the min-max normalized version $\widetilde{X}$ is defined as}
\begin{equation}
    \widetilde{X}
    \;=\;
    \frac{X - X_{\min}}{X_{\max} - X_{\min}},
    \quad
    \widetilde{X} \;\in\; [0,1].
\end{equation}
\rev{The normalized quantities $\widetilde{R}_v(\kappa_v), \widetilde{R}_c(\kappa_c), \widetilde{d}, \widetilde{\Delta d}$ are defined analogously.} The minima and maxima are determined from known bounds of the code distance $[d_{\min}, d_{\max}]$, and viable degree distributions for $R_v(\kappa_v), R_c(\kappa_c)$.

The reward is formulated as
\begin{equation}
    \widetilde{R}
    \;=\;
    \alpha \Bigl(
       \sum_{v \in V} \widetilde{R}_v(\kappa_v)
       \;+\;
       \sum_{c \in C} \widetilde{R}_c(\kappa_c)
    \Bigr)
    \;+\;
    \beta \,\widetilde{d}
    \;-\;
    \delta \,\widetilde{\Delta d},
    \label{eq:overall-reward}
\end{equation}
where \( V \) and \( C \) are variable and check nodes, \( \kappa_v \) and \( \kappa_c \) are their degrees, \( d \) is the code distance, and \( \Delta d = d_{\text{new}} - d_{\text{prev}} \) is the change in distance. The weights \( \alpha \), \( \beta \), and \( \delta \) balance degree reduction, distance preservation, and penalize distance reductions. The weights sum to 1 to ensure the reward is between 0 and 1.

To enforce \rev{weight constraints on parity checks}, we employ action masking:
\begin{equation}
    M(a|s) =
    \begin{cases}
        1, & \text{if action } a \text{ is permissible in state } s, \\
        0, & \text{otherwise},
    \end{cases}
\end{equation}
We exclude all adding operations from nodes with degree greater than the maximum desired weight, and all deleting operations from nodes with degree 1 or the minimum desired weight. Masked actions are assigned zero probability, and the remaining action probabilities are re-normalized to form a valid probability distribution. This preserves the theoretical properties of PPO, such as the trust region constraint and still ensures stable policy updates \cite{huang_closer_2022}. Most importantly, this allows us to restrict the agent to only focus on codes within the target weight, which enhances learning efficiency without compromising the algorithm's convergence guarantees. 
The mask can be adapted to codes with a variety of constraints on their structure. For example, \rev{the masked property for general stabilizer codes is symplectic orthogonality of \( H \), while
for CSS codes it is the orthogonality between \( H_X \) and \( H_Z \)}. \rev{We can also extend this to various important refined constraints are as needed, such as $k$-orthogonality which induces codes with transversal non-Clifford gates, and geometric locality or connectivity constraints that ubiquitously arise in physical and experimental scenarios.}

\begin{figure*}[ht]  
\centering
    \subfloat[\label{fig:reward}]{\includegraphics[width=0.32\textwidth]{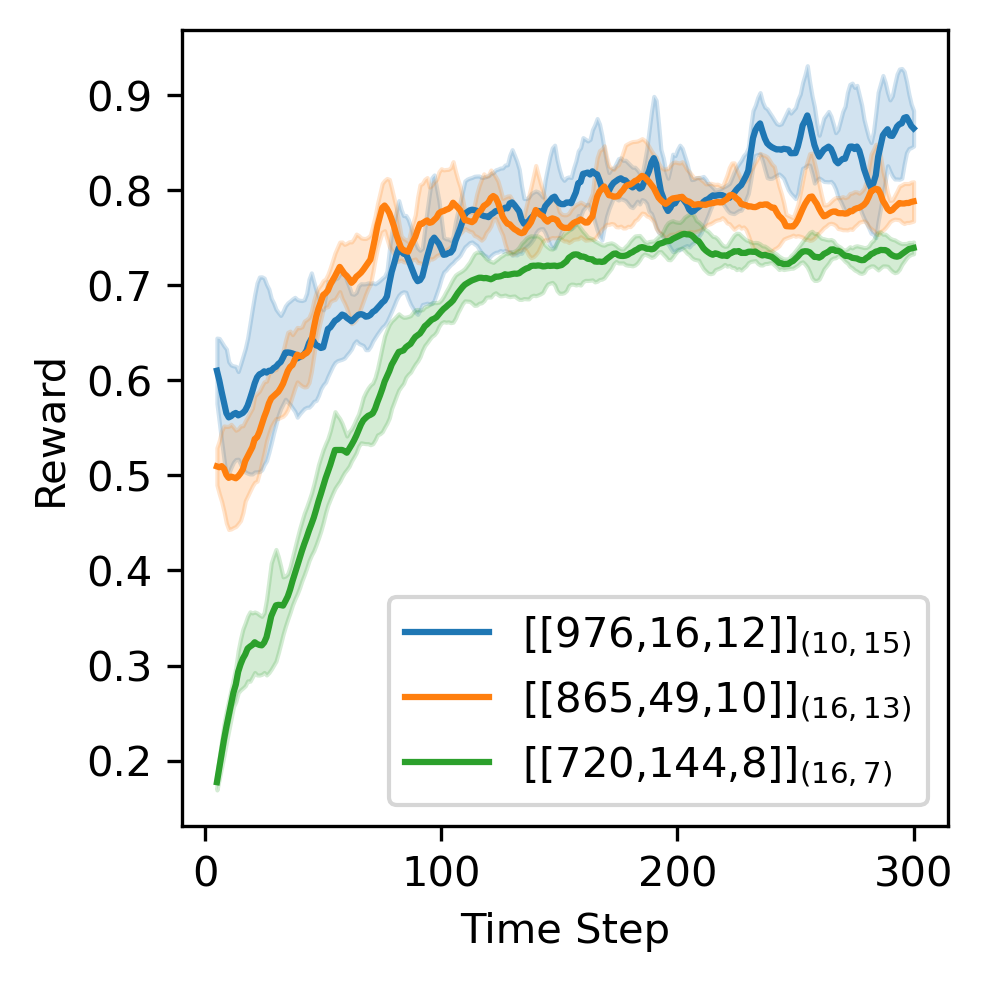}
    }
\hfill
    \subfloat[\label{fig:evolution}]{\includegraphics[width=0.32\textwidth]{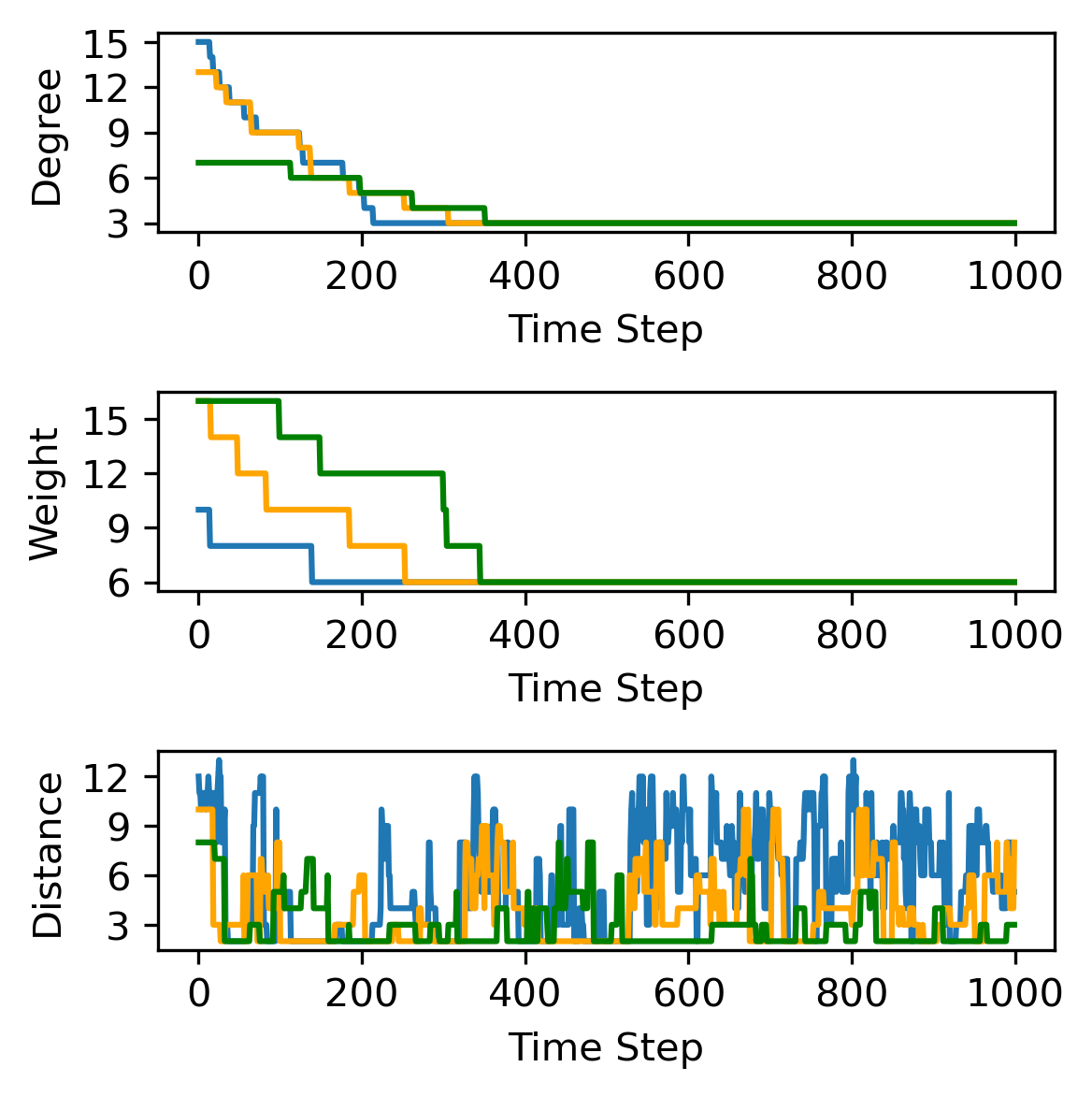}
    }
\hfill
   \subfloat[\label{fig:pca}]{ \includegraphics[width=0.32\textwidth]{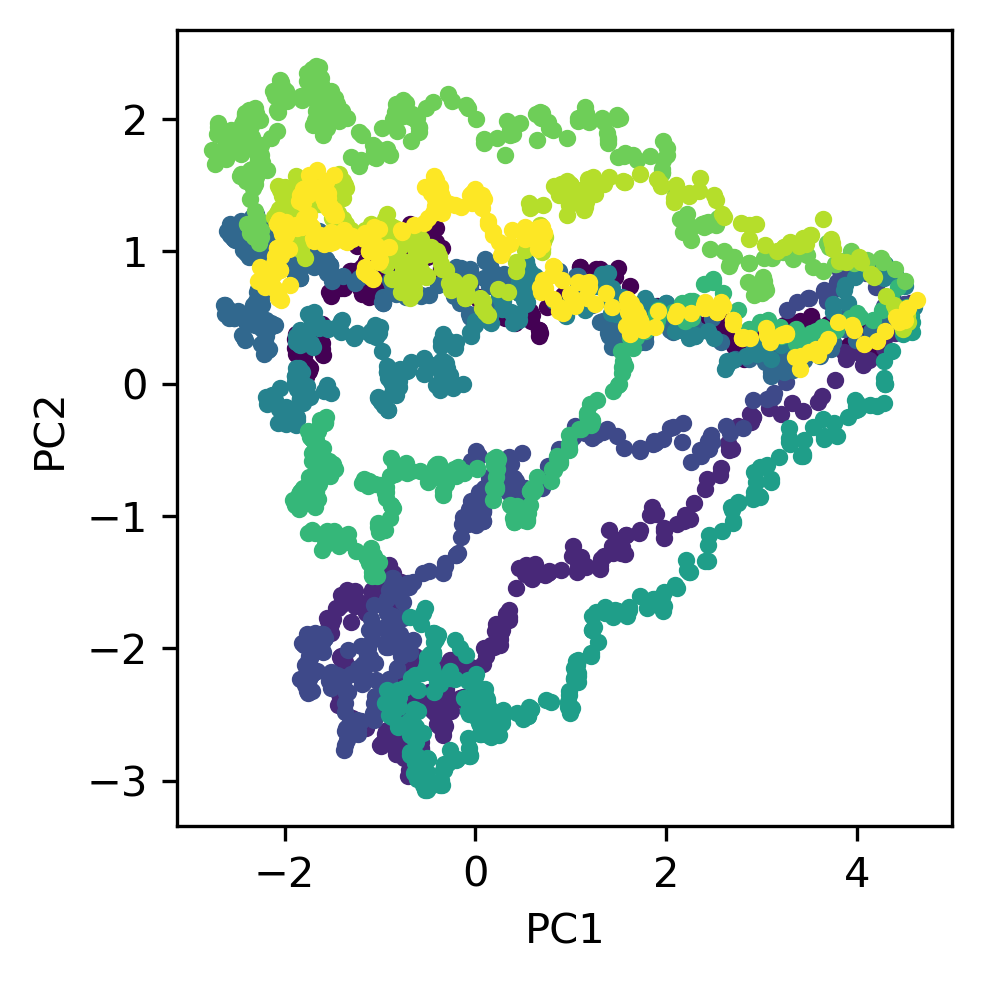}
    }

\caption{{Reinforcement learning-driven code design.} {(a)} Training trajectories of codes with varying parameters averaged over 3 runs. {(b)} Evolution of parameters in the three example codes throughout a single episode.
{(c)} Exploration of 10 episodes \rev{(represented by different colors)} over PCA decomposition of state space.}
\label{fig:rl-methods}
\end{figure*}

In Fig.~\ref{fig:reward}, we \rev{showcase} samples of \rev{reward} curves for our RL policy. The initial reward for codes with lower rate is higher, as there are more checks so low weight codewords are less likely to exist at random, while higher rate codes have less checks so low weight codewords are more likely to exist at random \cite{Lidar_Brun_2013}. All three displayed codes still converge to rewards in a similar range, although the learning process is more stochastic for codes with higher maximum distances. The evolution of weight, degree, and distance parameters over one episode of training on the \rev{three example codes are} shown in Fig.~\ref{fig:evolution}, and the area explored over 10 episodes in the state space of tanner graphs is shown in Fig.~\ref{fig:pca}. We see that as maximum degree and weight are gradually reduced, distance also decreases, then plateaus and rapidly increases. After this point both degree and weight remain in range and the policy begins to tweak distance. It is interesting that despite the $\Delta d$ term penalizing decreases in distance, the policy does not attempt to reduce the weight and degree in a more slow and careful way to perfectly preserve $d$ at each step. We provide two interpretations, either the policy is taking the path of least resistance and simply finds the problem easier to approach this way, or it could be learning from fluctuations in distance. This term both penalizes decreases and rewards increases in distance, so it may be pursuing the maximum possible reward in this setting, which is when a code with minimum $d$ code is modified into a code with maximum $d$ in one step, while the maximum weights and degrees are below 6 and 3. Regardless, the policy still produces codes that effectively minimize weight reduction overhead.


\section{Main Results}
\label{sec:results}

\rev{We now showcase the representative code discovery results achieved through our RL weight reduction scheme and discuss important comparisons with existing results. Additional information including more complete data and numerous auxiliary results and illustrations can be found in the Supplementary Information. }

\rev{We extend the standard $[\![n,k,d]\!]$ notation for QEC codes ($n,k$ are the number of physical and logical qubits respectively and $d$ is the code distance) to \qldpc{n}{k}{d}{w}{q}, where $w$ denotes the check weight and $q$ denotes qubit degree.}

\rev{\subsection{Overview of code discovery}}

\rev{Here we primarily focus on the following  setting: we apply our RL agent to hypergraph product codes, and aim to reduce to a maximum weight of six and qubit degree of three}, matching the parameters produced in Ref.~\cite{sabo_weight-reduced_2024}. These are favorable parameters for practical implementation and also allow us to make direct comparisons.
\rev{Note that our method can be adapted for any given weight or degree, and we also show examples for maximum weight eight and degree four.}
\rev{Furthermore, one can easily apply our method to general stabilizer codes settings and specialize to certain types including CSS codes, product codes, $k$-orthogonal codes and so on as needed, as addressed in Sec.~\ref{sec:methods} by adapting the action masking logic accordingly.}


\rev{Hypergraph product codes~\cite{tillich_quantum_2009} provide an elegant framework for constructing quantum codes from classical ones and serve as a prototypical model in the study of qLDPC codes with desirable code parameters and FT properties~\cite{tillich_quantum_2009,krishna_fault-tolerant_2021, connolly_fast_2024,breuckmann_qldpc_2021}}. \rev{More explicitly, given parity check matrices of classical codes $H_1$ and  $H_2$, define
\begin{align}
    H_X &= \begin{pmatrix}
            H_1 \otimes I_{n_2} & I_{r_1} \otimes H_2^T
          \end{pmatrix}, \label{eq:H_X} \\
    H_Z &= \begin{pmatrix}
            I_{n_1} \otimes H_2 & H_1^T \otimes I_{r_2}
          \end{pmatrix}, \label{eq:H_Z}
\end{align}
represent the $X$ and $Z$ check matrices of a CSS code respectively.
Here \( H_1 \) and \( H_2 \) are the {check} matrices of the original classical codes, and \( I \) denotes identity matrices of appropriate dimensions. This construction ensures orthogonality between \( H_X \) and \( H_Z \) for any pair of linear classical codes and therefore produces a valid CSS stabilizer code.} In this work we use the same classical code for $H_1$ and $H_2$.

\rev{We generate new codes by executing our RL scheme on} hypergraph product base codes constructed from all classical codes with $n\leq 30$ from the best known linear codes database in the GUAVA package in GAP \cite{GUAVA, GAP4}, which we collectively refer to as HGP-30. \rev{Several examples of codes with particularly large distances beyond this range will also be included.} All results were produced using an i7-13700HX CPU and RTX-4060 GPU. With more extensive training it is feasible to further optimize code parameters or scale to larger sized codes \cite{hilton_scaling_2023}.

\begin{figure}[ht]
    \centering
    \includegraphics[width=1\linewidth]    {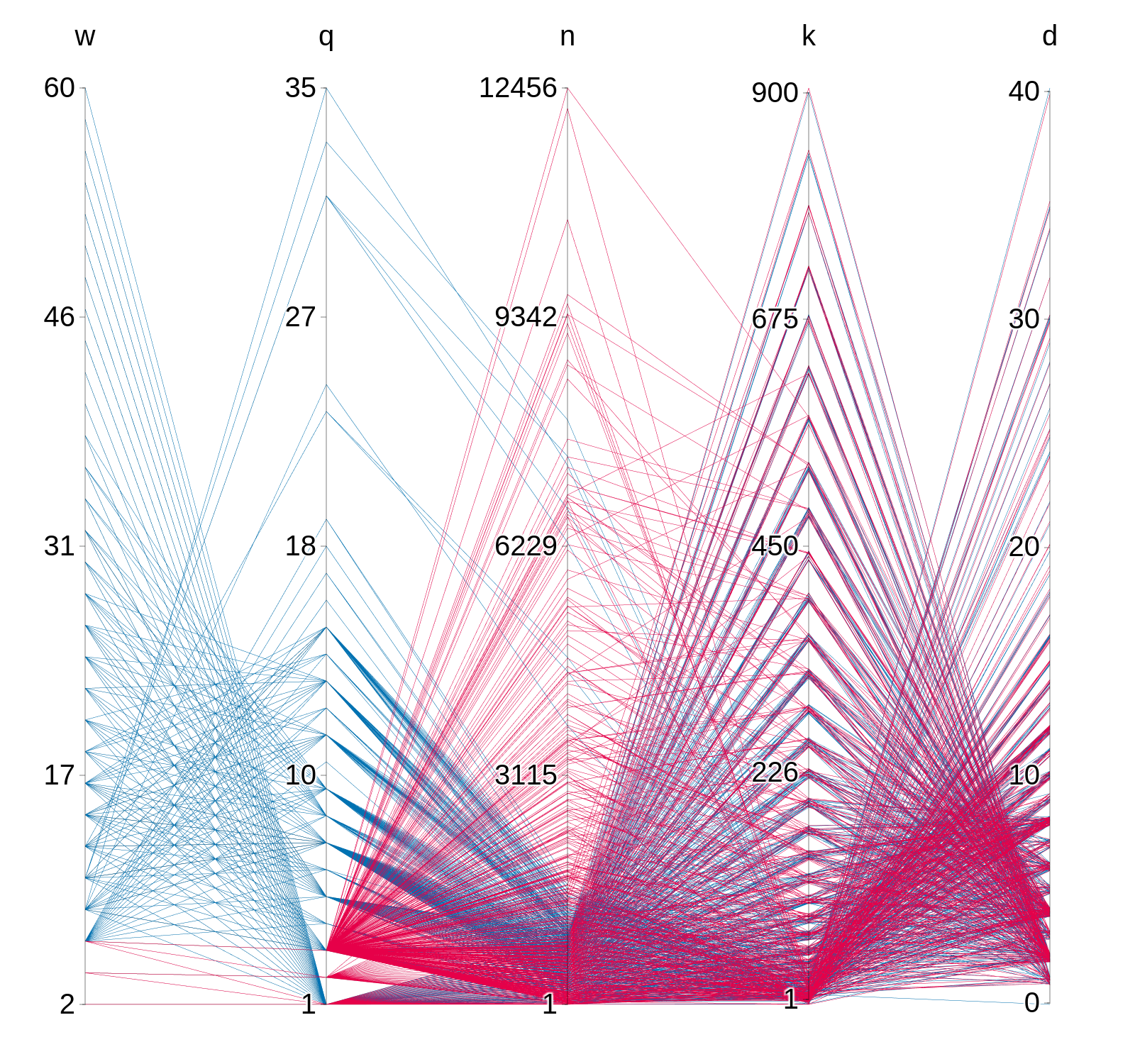}

    \caption{ {Parallel coordinates plot comparing hypergraph product base codes (blue) and RL-optimized codes (red) after weight reduction.} For each color, {475 codes} \rev{(including 10 high-distance ones beyond the HGP-30 regime)}   with varying parameters are shown. Each vertical axis is normalized to the maximum observed value for that parameter, and each line traces a single code’s parameters across all axes. }
    \label{fig:parallel_coords}
\end{figure}

\rev{We visualize the code parameter combinations of the base and new codes produced by our RL-based weight reduction scheme with a parallel coordinates plot -- see Fig.~\ref{fig:parallel_coords}.
This illustrates how our RL-based weight reduction scheme modifies code parameters. Initially, the $w$ and $q$ parameters can be quite large, and while most codes have relatively balanced weight and degree, there exist a number of codes which are highly skewed. After our RL-based weight reduction, $n$ tends to increase while $w$ and $q$ decrease, and the $k$ and $d$ values are unchanged. Unsurprisingly, $k$ and $d$ exhibit an inverse relation, as codes with higher $k$ tend to have lower $d$, and vice versa.}
\rev{Our method uncovers a broad spectrum of new low-weight and degree codes with considerably better code parameters compared to known constructions, as we will demonstrate explicitly through suitable comparisons now.}

\begin{figure}
    \centering
    \includegraphics[width=1\linewidth]{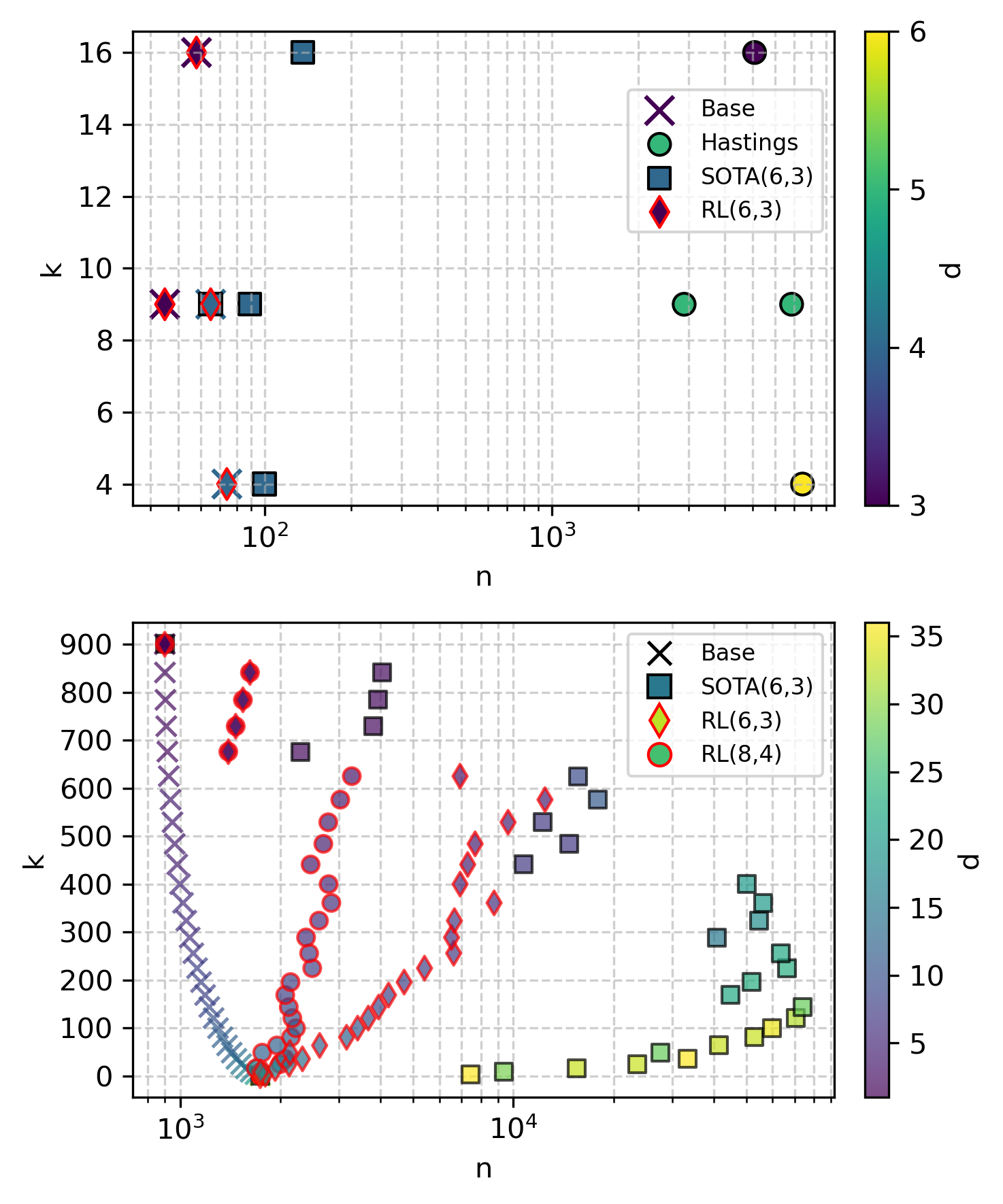}
    \caption{ {Comparisons of codes discovered by our RL-based scheme and existing weight reduction methods.} {(top)} Comparison with Hastings~\cite{hastings_quantum_2023} (data taken from Ref.~\cite{sabo_weight-reduced_2024}) and SOTA results from Sabo et al.~\cite{sabo_weight-reduced_2024}. {(bottom)} Comparisons with SOTA results on all hypergraph product codes constructed from $n\leq30$ classical codes. Explicit code parameters are shown in Table~\ref{tab:qldpc-subscript},~\ref{tab:qldpc-comparison}.}
    \label{fig:hastingsandn30}
\end{figure}

\subsection{Weight reduction comparisons}

\begin{figure*}[t]
    \centering
    
     \subfloat[\label{fig:overheadoverall}]{\includegraphics[width=0.32\textwidth]{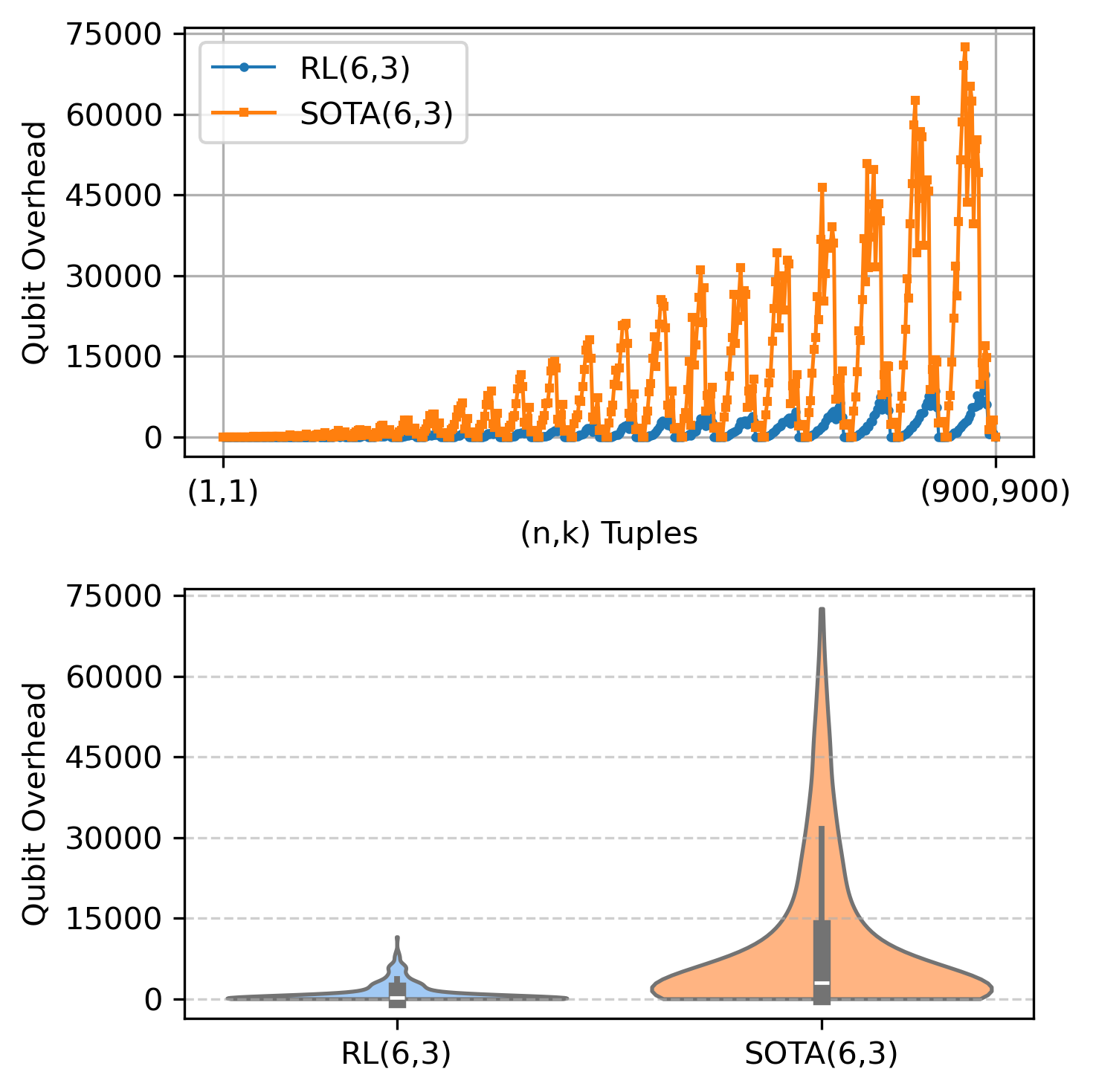}
    }
    \hfill
    \subfloat[\label{fig:ratevsreld}]{\includegraphics[width=0.32\textwidth]{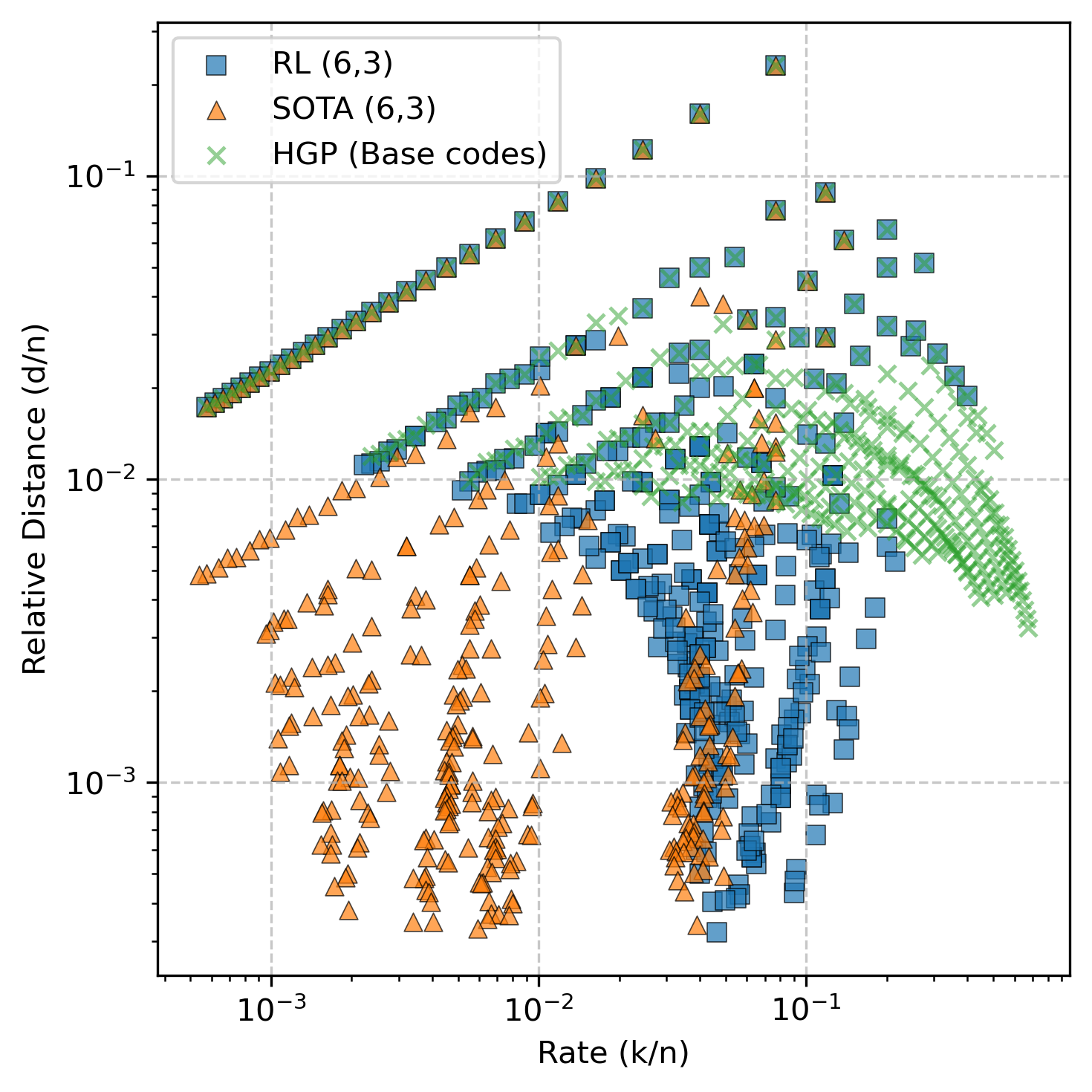}
    }
    \hfill
    \subfloat[\label{fig:rldistances}]{\includegraphics[width=0.32\textwidth]{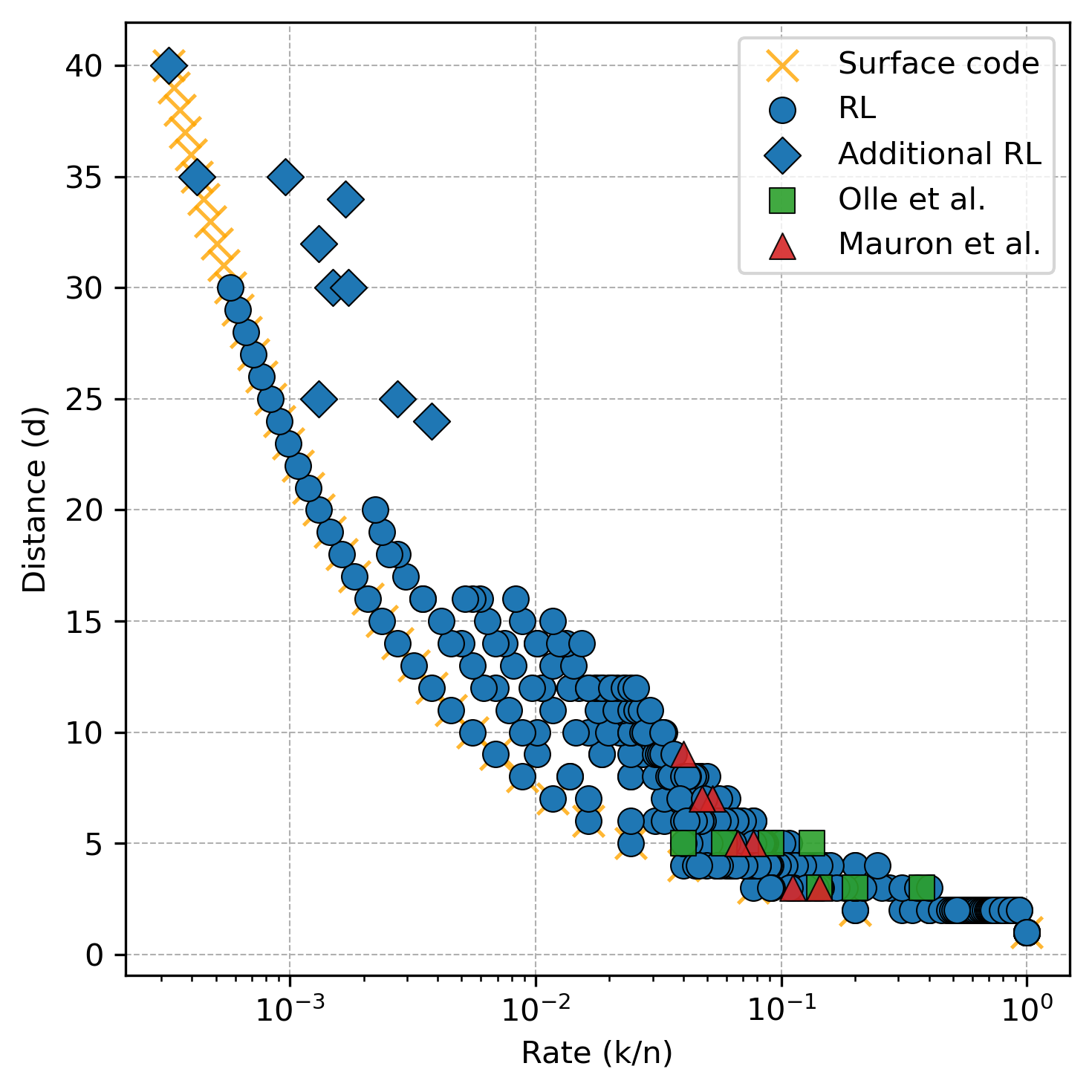}
    }

    \caption{ {Comparisons against SOTA on weight reduction and RL code design.}  
    {(a)}~Comparison of qubit overheads of weight reduction to (6,3) between RL and SOTA on all hypergraph product codes constructed from classical codes with $n\leq30$.
    {(b)}~Rate vs.\ relative distance for weight-reduced vs.\ hypergraph product base codes.
     {(c)}~Comparisons of code parameters against various alternative RL methods for code design. \rev{Ten additional (6,3) codes (labeled by diamonds) beyond the HGP-30 regime were produced using our RL model to display examples of particularly high distance codes.}}
    \label{fig:distance-overhead-comparisons}
\end{figure*}

\rev{We first compare the effectiveness of our RL-based method and existing weight reduction methods.}
\rev{In Fig.~\ref{fig:hastingsandn30}, we illustrate the direct comparisons between the parameters of codes discovered by our RL policy and earlier weight reduction methods. In the low-rate and low-distance regime, Hastings’ early method \cite{hastings_quantum_2023} require thousands of qubits. The state-of-the-art (labeled by SOTA) results from Sabo et al. \cite{sabo_weight-reduced_2024} lower the qubit overhead by more than one order of magnitude compared to Ref.~\cite{hastings_quantum_2023} but some overhead remains, while our RL method does not need overhead at all in this regime. For larger codes, the overhead of our method exhibits roughly 1--2 orders of magnitude lower overhead on (6,3) codes (see also Fig.~\ref{fig:distance-overhead-comparisons} for further information).} 
\rev{We also compare the overheads for codes reduced to weight and degree (8,4), which are significantly smaller at moderate to high rate codes. The qubit overhead for (6,3) and (8,4) are not related in a straightforward way, for example, the qubit overhead when reducing to (6,3) is increasing almost monotonously at ($0<k<600$), while for (8,4) it shows decreasing trends at many ranges of $k$ values ($0<k<200$), ($200<k<300$), ($350<k<450$). This provides a glimpse into the trade-off of $w$, $q$ against $n$, $k$, $d$, although a more thorough exploration of different $w,q$ combinations would be needed to fully map these trade-offs. These points are not necessarily optimal, and the inefficiencies are unknown, so the true relationship between the upper bounds on $n,k,d,w,q$ could also turn out to be vastly different.} 

\begin{figure*}[t]
    \centering
    
    \adjustbox{max width=0.32\textwidth}{%
        \includegraphics{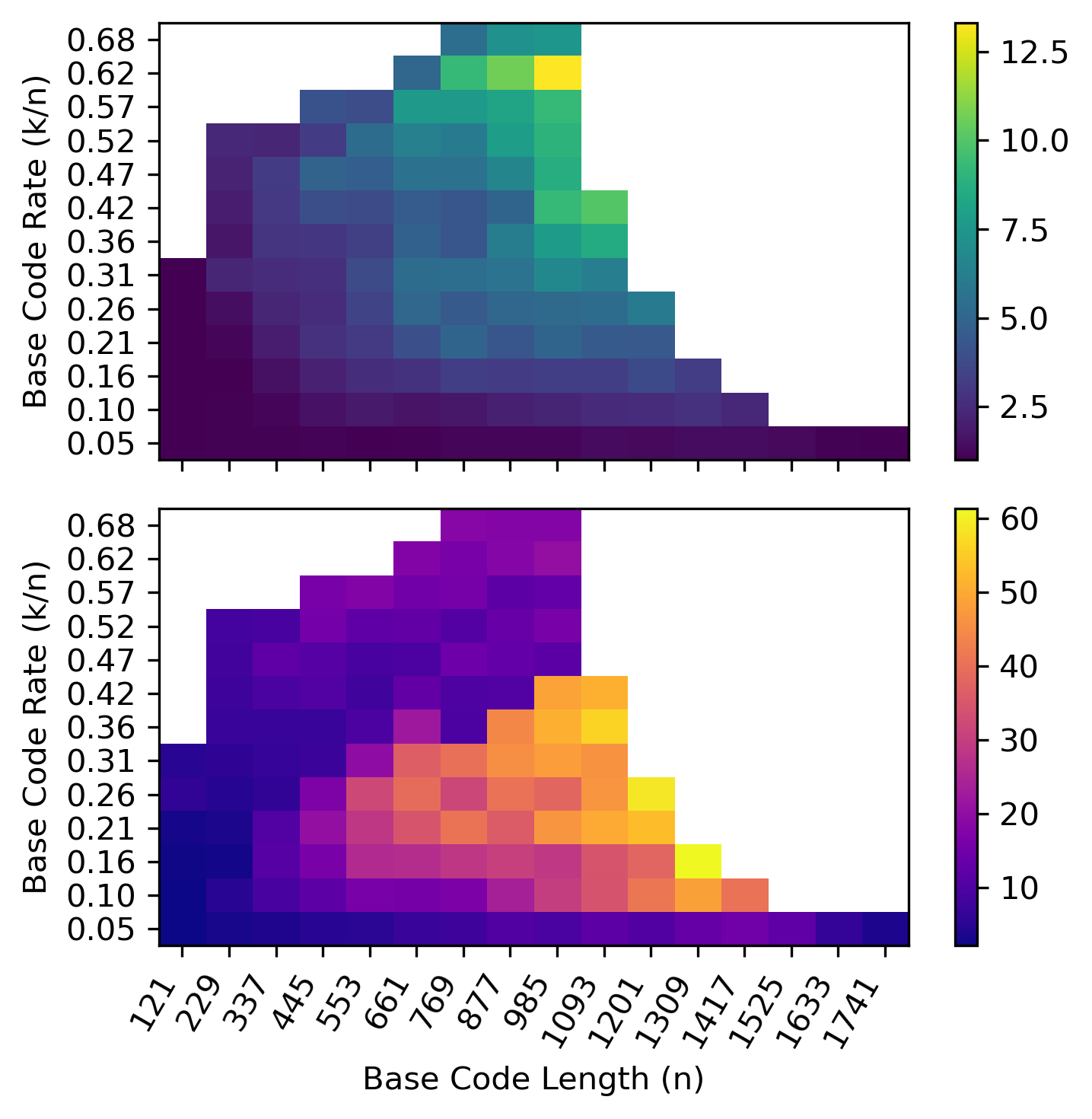}
    }
    \hfill
    \adjustbox{max width=0.32\textwidth}{%
        \includegraphics{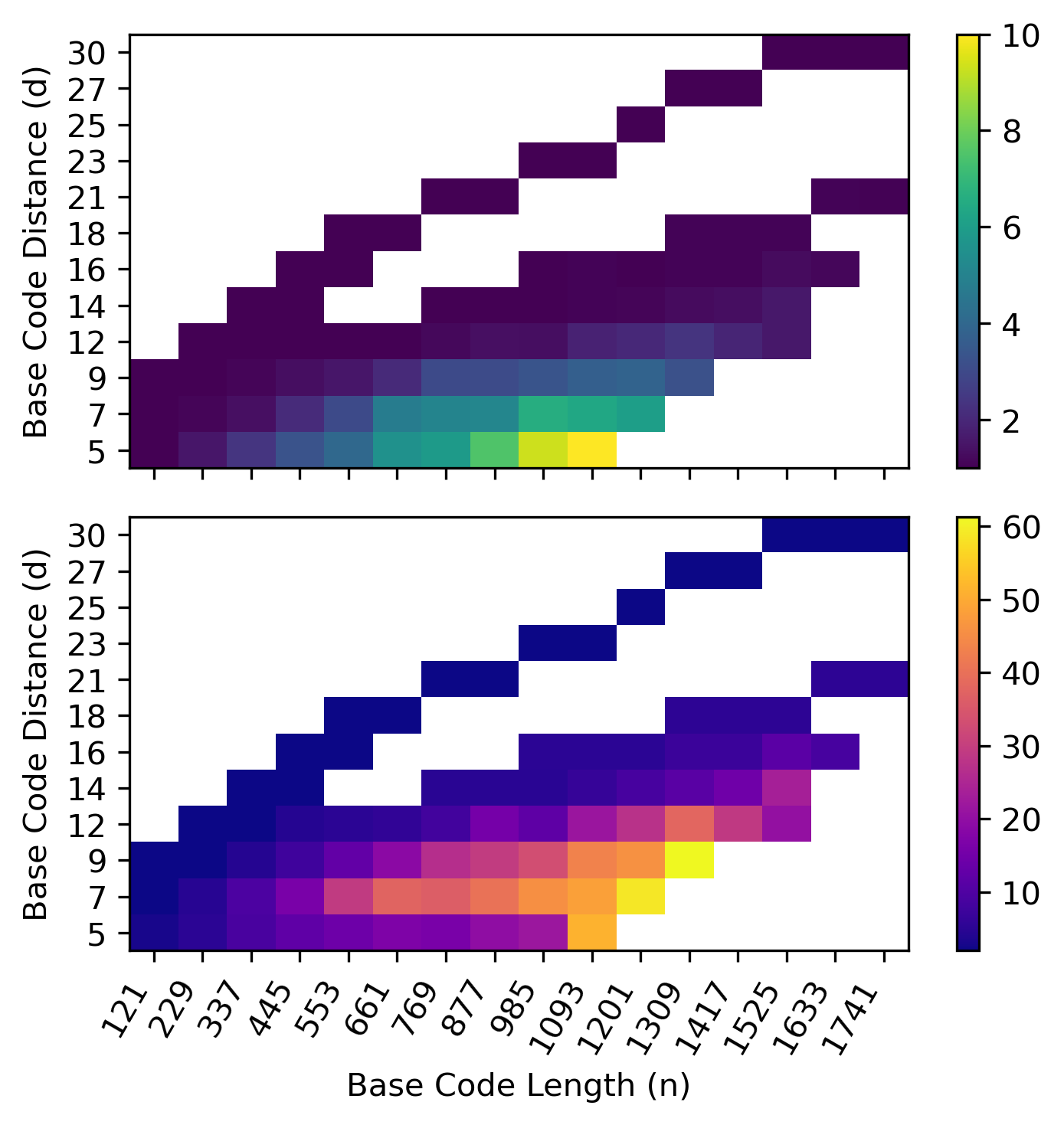}
    }
    \hfill
    \adjustbox{max width=0.32\textwidth}{%
        \includegraphics{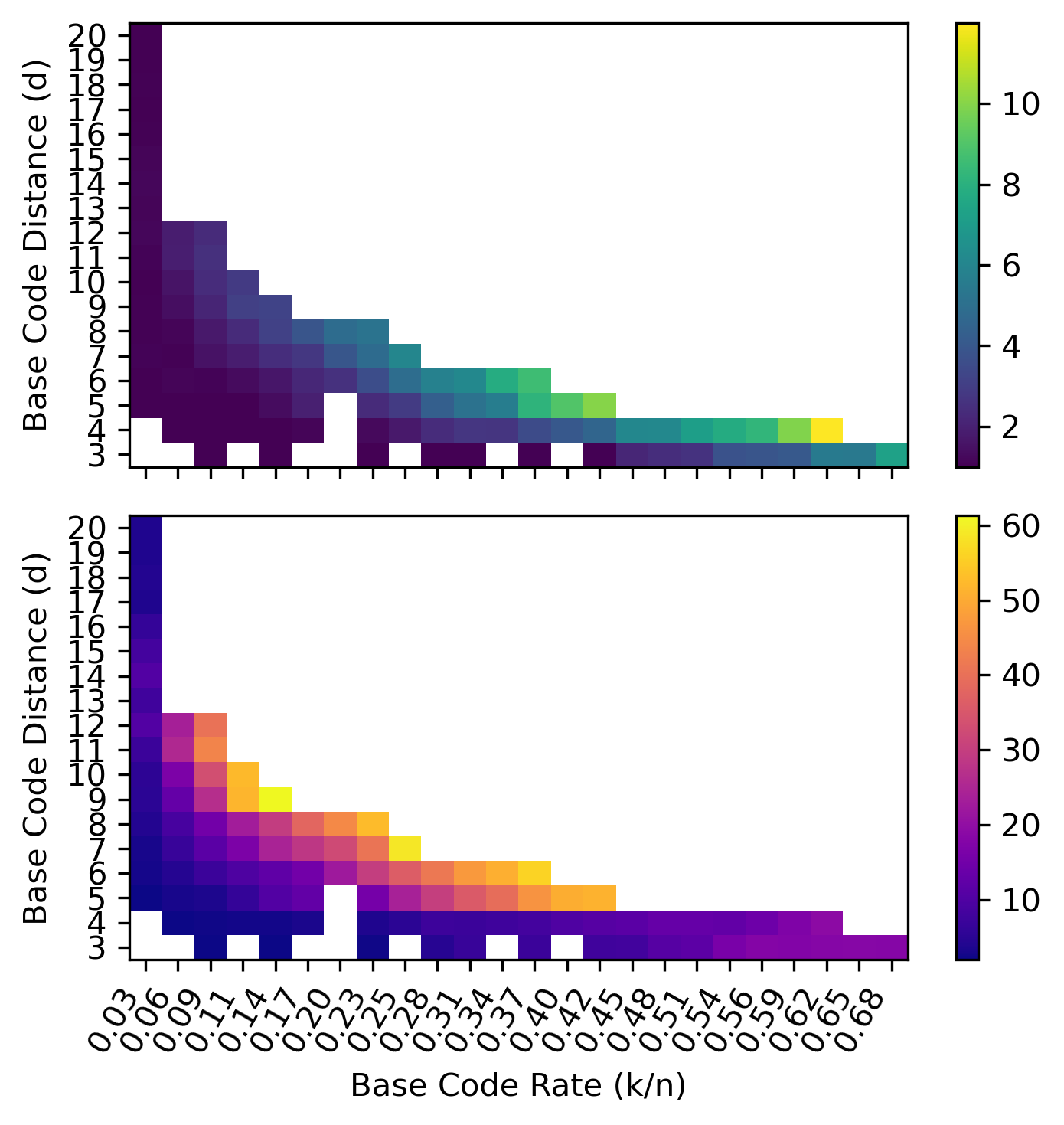}
    }

    \caption{{Breakdown of overhead factors \rev{shown by  heatmaps} at varying $n$, $k$, $d$ parameters.} \rev{The top and bottom rows correspond to codes discovered by our RL weight-reduction scheme and Sabo et al.'s method, respectively.}
    Gradients are binned for ease of visualization and not exact representations of overhead factors, as seen in the varying scales.}
    \label{fig:overhead-comparisons}
\end{figure*}

\rev{Then in Fig.~\ref{fig:overhead-comparisons} we depict the overhead factors (the ratio of the qubit number after and before weight reduction) for various combinations of code parameters, highlighting the clear advantages in efficiency of our RL-based method.}
As shown, the overhead of RL weight reduced codes tends to gradually increase with code rate $k/n$, while for Ref.~\cite{sabo_weight-reduced_2024}, the overhead peaks around moderate rates ($0.1<k/n<0.40$) and sharply decreases after this point. Our RL method exhibits lower qubit overhead for all $465$ codes considered, with the greatest difference at low to moderate rates. The maximum observed reduction in qubit overhead among the examples is about 73x for a \qldpc{1109}{9}{14}{8}{13}  code. 
\rev{Further analysis shows that our RL agent tends to have the highest overhead on $d=4$ codes, and the overhead factor decreases as distance increases (mostly because this means $k/n$ is decreasing). We also note that the areas with highest overhead factor locally tend to be around the frontier of $k/n$ vs $d$, which is evident from Fig.~\ref{fig:overhead-comparisons} and Fig.~\ref{fig:paretofronts} in Supplementary Information. Overall, this suggests that for our RL agent, the bottleneck primarily arises from $k/n$ instead of $d$, which bodes well for the performance of our RL agent at larger distances and code sizes.}


Overall, our RL agent significantly reduces the qubit overhead needed to achieve a target weight and degree, including codes relevant in the near-term parameter regime. To exemplify the effectiveness of our RL policy, in Fig.~\ref{fig:overheadoverall} we show our RL policy tends to require an overhead of $10^3$ to $10^4$ physical qubits, whereas Ref.~\cite{sabo_weight-reduced_2024} requires an overhead of $10^4$ to $10^5$ physical qubits, \rev{representing 1--2 orders of magnitude's improvement over SOTA}. \rev{In particular, our method exhibits vastly smaller overhead at moderate to large code sizes.}  This is especially clear in an example beginning with a \qldpc{2500}{100}{16}{29}{19} base code: our RL agent produces a \qldpc{6100}{100}{16}{6}{3} code, which is within reach for near-term hardware \cite{preskill_beyond_nodate}, whereas the SOTA method in Ref.~\cite{sabo_weight-reduced_2024} yields a \qldpc{144772}{100}{55}{6}{3} code with a substantially more severe qubit count requirement. \rev{Notice that despite the large overhead, the approach of Ref.~\cite{sabo_weight-reduced_2024} can also increase distance. This may potentially produce codes interesting in their own right, although up to orders of magnitude larger than the original code. The codes discovered by our RL agent tend to have better overhead to distance scaling as shown in Fig.~\ref{fig:ratevsreld}. }
For example, our RL agent finds a \qldpc{2257}{25}{15}{6}{3} code from a \qldpc{1525}{25}{15}{12}{15} base code, while for the same $n$ and $k$, the method given in Ref.~\cite{sabo_weight-reduced_2024} produces a \qldpc{2257}{25}{13}{6}{3} code from a \qldpc{277}{25}{6}{6}{10} base code. Other codes with similar $k$ and $d$ produced by Ref.~\cite{sabo_weight-reduced_2024} require even larger overhead to obtain comparable $d$ and are shown in Table~\ref{tab:rl-sabo-comparison}. \rev{This comparison may be conservative as the method given in Ref.~\cite{sabo_weight-reduced_2024} can produce vastly different results for base codes with the same parameters, and even permutations of the same code, while our RL agent is running on relatively modest computational resources.}
\rev{To conclude, the codes discovered by our method are expected to exhibit even larger advantages at larger distances, although at the price of increasing computational demands \cite{hilton_scaling_2023}.}

\subsection{RL code design comparisons}

Check weight constraints have scarcely been considered in previous studies applying numerical methods to code design, especially for RL, largely because they have focused on codes with few physical qubits, or with relatively constrained state spaces.
\rev{Here we make specific comparisons with the most recent works by Olle et al.~\cite{olle_simultaneous_2024} and Mauron et al.~\cite{Mauron_2024}.}
When making the comparisons, we still apply the restriction of maximum weight 6 and degree 3, although we are able to achieve even better parameters if we relax these constraints, at the price of the code performing worse in practice.

Olle et al.~\cite{olle_simultaneous_2024}  find codes up to $d\leq5$ and $n\leq25$. A circuit level representation and Knill-Laflamme condition based reward function is used, and requires tracking \rev{all error operators with Hamming weight equal to the target distance}. This approach circumvents the sparsity of a distance reward, but is exponentially memory intensive, and they create a road map which requires a GPU with approximately 80GB of VRAM to find codes with $d=10$. Mauron et al.~\cite{Mauron_2024} find codes with up to $d\leq9$ and $n<25$. A tensor network representation is used with a distance-based reward function which is not restricted by memory requirements, \rev{making it possible to design codes with larger $d$, but still were unable to surpass $d=10$}. 

\rev{We compare the codes discovered by our RL agent with those from Refs.~\cite{olle_simultaneous_2024,Mauron_2024} along with surface code in  Fig.~\ref{fig:rldistances}.} \rev{All methods can produce codes that outperform the surface code in terms of rate and distance}. 
\rev{While other methods are restricted to $d\leq9$, our RL scheme can design codes in a significantly larger distance regime, discovering codes up to $d=40$ and ones outperforming the surface code with $d = 35$.} \rev{Note that we did not run a comprehensive sweep outside the HGP-30 regime} due to computational constraints. \rev{Although the rates are comparable at small code lengths, our RL agent discovers codes with a maximum number of logical qubits $k=900$, while previous works have been limited to codes with single digit $k$}. 
Also, these are the first RL designed codes with parameters large enough to be relevant to devices in the approaching Megaquop regime~\cite{preskill_beyond_nodate}, where it is widely believed that relatively high distance codes with at least \rev{ $d\gtrapprox 20$} will be necessary to reach logical error rates on the order of $10^{-6}$ \cite{acharya_quantum_2024, xu_constant-overhead_2024,preskill_beyond_nodate}.


\begin{figure*}[t]
    \centering
    \subfloat[\label{fig:hgpscatterplots}]{ \includegraphics[width=0.48\textwidth]{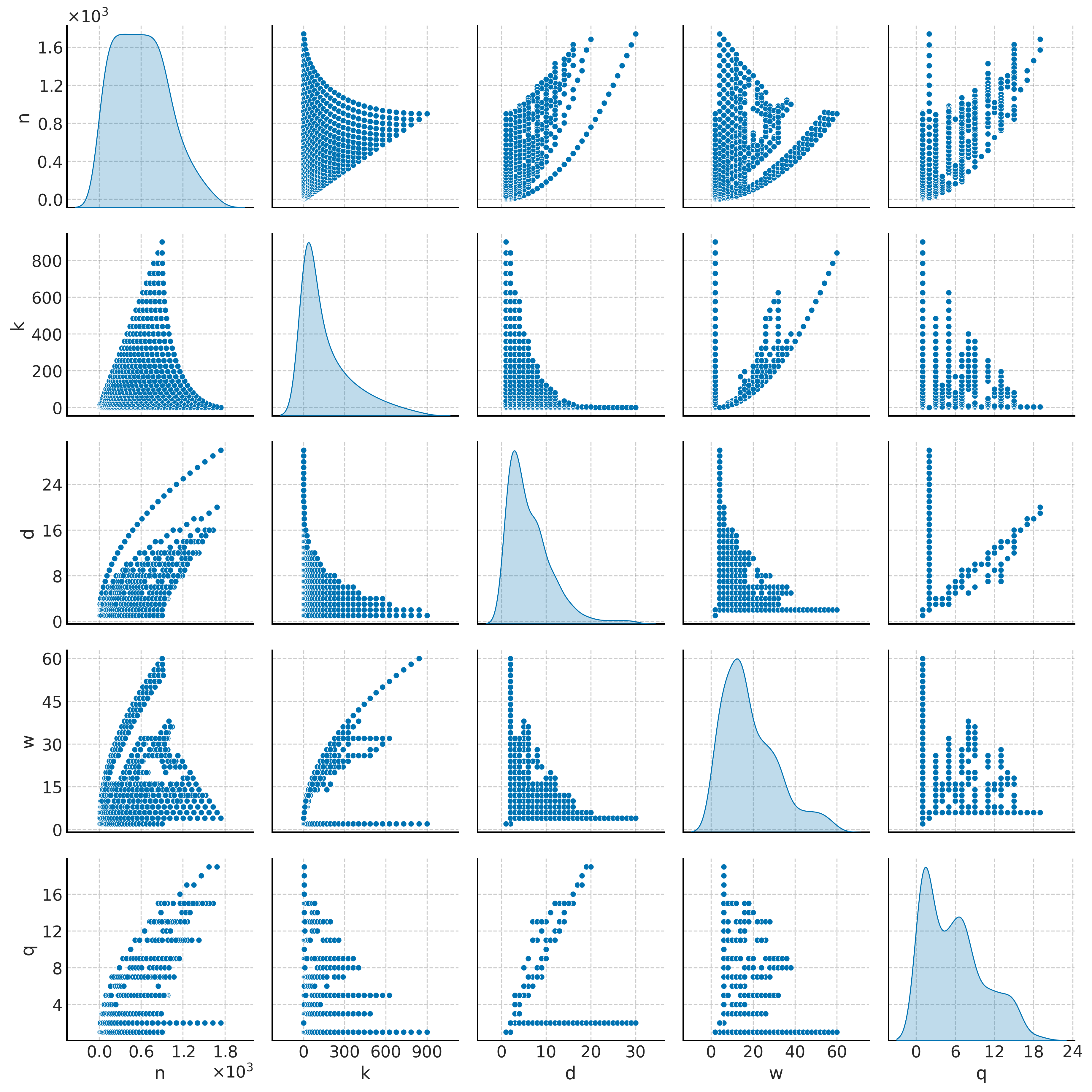}
    }
    \hfill
    \subfloat[\label{fig:rlscatterplots}]{ \includegraphics[width=0.48\textwidth]{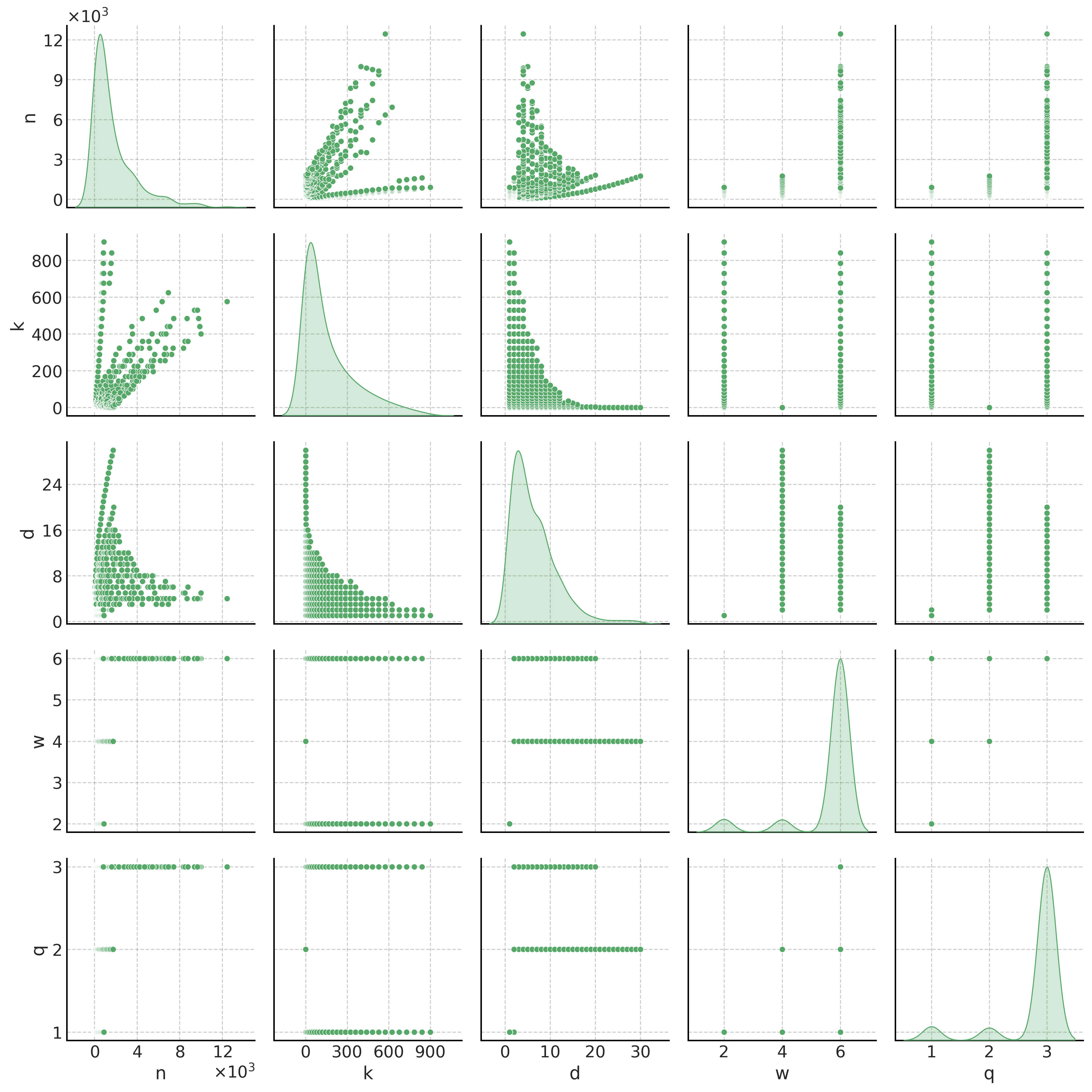}
    }
    \caption{
    {Pairwise scatter plots for the parameters of the HGP-30 base codes and RL codes.}
    {(a)} Hypergraph product code with parameters $n,k,d,w,q$ before weight reduction,
    {(b)} RL codes after weight reduction.
    Each subplot compares two parameters (off-diagonal) while the diagonal entries depict individual parameter distributions.
    }
    \label{fig:combined_scatterplots}
\end{figure*}

\subsection{\rev{Additional discussions and results}}

\rev{In Fig.~\ref{fig:hgpscatterplots}, we present pairwise scatter plots for the parameters of the base hypergraph product codes and RL weight reduced codes, which provide abundant insights into the behavior of the RL scheme and codes.}
\rev{ Prior to weight reduction, the $w$ and $q$ values correlate with $n,k,d$, while the correlations vanish after our RL agent applies weight reduction and $w$ and $q$ become constant as seen in Fig.~\ref{fig:rlscatterplots}. The data points for $d$ and $n$ make curves that are well described by $d=O(\sqrt{n})$ for each $k$ value, as shown in Fig.~\ref{fig:metaanalysisd} and Fig.~\ref{fig:metaanalysisk} in Supplementary Information. This confirms that finite-size scaling aligns with known asymptotic bounds on hypergraph product codes~\cite{tillich_quantum_2009}, and demonstrates more fine-grained details of $d$ vs.~$n$ scaling and the effects of weight reduction.} 


\begin{figure}[t]
    \centering
    \includegraphics[width=0.8\linewidth]{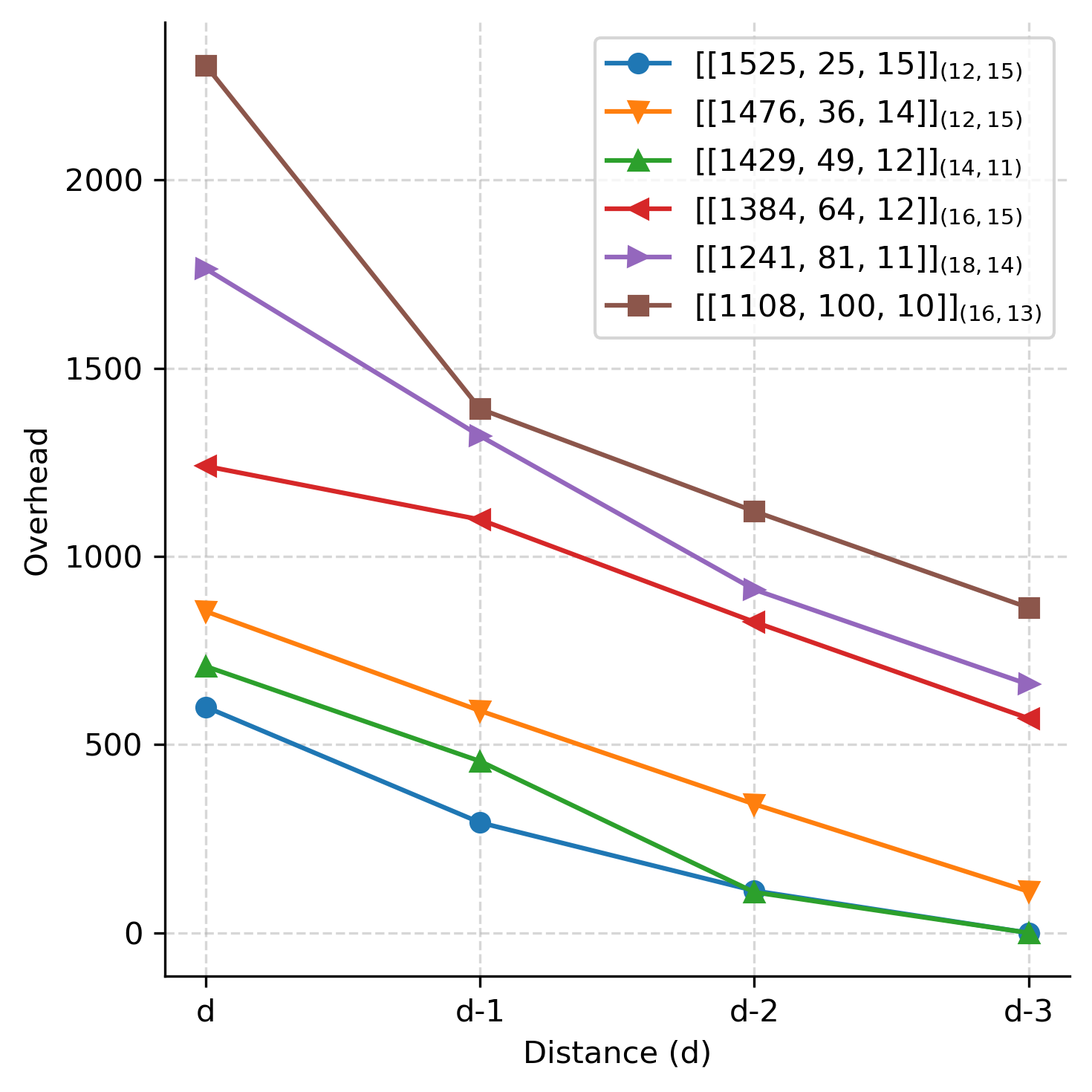}
    \caption{\rev{{Minimum qubit overheads under relaxed distance constraints.} Permitting the distance $d$ to decrease can reduce the total qubit overhead.} Code parameters are shown in Table~\ref{tab:qldpc_overheads}}
    \label{fig:reducedistance}
\end{figure}

\rev{It is also natural to ask how the behaviors of our scheme change if some loss of distance is allowed, which is also considered in the original formulation of weight reduction in Ref.~\cite{hastings_weight_2016}. In Fig.~\ref{fig:reducedistance}, we showcase 6 examples with varying $k$ and $d$ values to demonstrate how overhead can be further reduced at the cost of a loss in distance. The required overhead tends to increase with rate, although the \qldpc{1429}{49}{12}{14}{11} and \qldpc{1476}{36}{14}{12}{15} codes are an exception, possibly because the difference in $d$ is 2, instead of 1 as in other codes, and thus more influential on the overhead. This further hints at the trade-off between $n$, $k$, $d$ at finite sizes, and may be of interest for near-term experiments when there are hard constraints on $n$. Also, it is possible that applying weight reduction to a larger code while allowing a small decrease in distance may yield better parameters than directly applying to a code with the target parameters.}


On a last note, the creation of linearly dependent stabilizers \rev{(corresponding to meta-checks)} when using product constructions \rev{has been overlooked in the context of weight reduction}. In both methods we consider $k$ to be fixed. However, after removing linearly dependent rows, $k$ can be increased by a multiplicative factor without affecting other code parameters~\cite{ostrev_classical_2024}. This applies to all the product codes considered, and our weight reduction procedure does not tend to prevent the formation of resulting linear dependencies. For instance,  the \qldpc{6100}{100}{16}{6}{3} code discovered by our RL policy has true rank 5683 and thus it actually has parameters \qldpc{6100}{417}{16}{6}{3}. 
\rev{This was not directly factored into our reward due to computational constraints as it slows down training and is mostly a feature of product code constructions that does not apply to weight reduction or code design in general, but the true $k$ is generically even better if we take meta-checks into account.}

\section{Conclusion and outlook}
\label{sec:conclusion}

\rev{In this work, we introduced a powerful new scheme for designing low-weight stabilizer QEC codes based on a highly efficient RL-based algorithm for weight reduction,
which takes the effective route of starting with a code with the target distance and then optimizing weight, rather than designing a code from scratch.}
\rev{This method enables the discovery of an abundance of new low-weight codes 
that extend well beyond the previously accessible code parameter regime even with modest computational resources. 
In particular, while existing numerical methods generally stagnate at single-digit distance codes, our approach systematically generates efficient low-weight codes with distances in the tens---a regime expected to be crucial for experimental developments in the coming years.
As low-weight QEC codes are critical components of fault-tolerant quantum computing, our findings pave the way for more feasible implementation of high-performance QEC, potentially bringing fault tolerance closer to realization in the near future.
}

\rev{Moreover, from the machine learning perspective, we have demonstrated that RL is particularly well-suited for stabilizer code design problems and, notably, far more scalable than previously thought. }
\rev{
Our simple model is able to circumvent previous obstacles in designing codes with high distance and low weight, and exhibit great potential for further scalability, as discussed.}
\rev{
Our current results are produced by running on 16 cores, and with about a thousand cores we expect to be able to design product codes with a few million qubits and non-product codes with a few thousand qubits. 

The main bottleneck in these estimates is the distance calculation. An avenue for future work could be using the spectral gap as part of the reward, which is commonly used to reason about distance in expander based constructions (although mostly in the asymptotic setting) 
The spectral gap is computable in polynomial time, as opposed to distance which is NP-hard \cite{kapshikar_hardness_2023}, and highly parallelizable. This is a good heuristic to identify codes likely to have high distance, but does not fully replace the distance calculation. With further tuning we expect the efficiency of the algorithm and thus the accessible code sizes to improve further. }


\rev{It should be noted that different choices of reward functions, model architectures, learning algorithms, hyper-parameters, and learning representations, etc., often lead to vastly different outcomes \cite{henderson_deep_2019}. It is likely that  further tuning to this framework and scaling up computational resources \cite{hilton_scaling_2023} can lead to improvements of the results.
Furthermore, as noted in Sec.~\ref{sec:methods}, our RL model can be adapted to codes with a variety of constraints on their structure, including geometry constraints and logical gates or symmetries, which would be worthy of further study.} 
\rev{In particular, previous RL code design has focused solely on the memory perspective. As a key challenge for fault tolerance, it is crucial to address the discovery of codes with desired logical operations (Ref.~\cite{Chen_2022} has studied the identification of logical operations for a given code using machine learning).}
\rev{Additionally, this framework can naturally be extended to design qudit codes, by modifying the action space to take an additional parameter.}
\rev{Other future directions we consider particularly important include investigating the compatibility of our high-performance RL-designed codes with different experimental platforms and architectures, further analyzing their fault tolerance properties, and establishing a more complete understanding of the tradeoffs between distance, weight and other code parameters.}

\rev{To conclude, our work highlights a promising new avenue where artificial intelligence can advance quantum computing through QEC code design. 
We anticipate vast opportunities for future work using artificial intelligence to discover codes and fault tolerance strategies at finite sizes, which hold great promise in uncovering new constructions that far surpass human-designed ones and accelerate the realization of scalable quantum technologies.
}

\begin{acknowledgments}
 ZWL is supported in part by a startup funding from YMSC, Tsinghua University, and NSFC under Grant No.~12475023.
\end{acknowledgments}

\bibliography{main-250118}

\widetext
\newpage
\section*{Supplementary Information}

The Supplementary Information includes additional data and diagrams that support and enrich the findings presented in the main text.

\subsection{Data tables}

\begin{table}[ht]
\centering
\caption{\rev{Some examples of Comparison of Hastings, Sabo et al.~\cite{sabo_weight-reduced_2024}, and our RL weight reduction methods. Results using Hastings' method are obtained from Ref.~\cite{sabo_weight-reduced_2024}.} Data is also shown in Fig.~\ref{fig:hastingsandn30} } 
\label{tab:qldpc-subscript}
\begin{tabular}{l|lla}
\hline
\hline
{Base Code} & {Hastings \cite{hastings_weight_2016}} & {Sabo et al.~\cite{sabo_weight-reduced_2024}} & {RL(6,3)}\\
\hline
\qldpc{45}{9}{3}{8}{3}
  & \qldpc{2892}{9}{5}{6}{6}
  & \qldpc{65}{9}{4}{6}{3}
  & {\qldpc{45}{9}{3}{6}{3}}\\

\qldpc{74}{4}{4}{6}{4}
  & \qldpc{7466}{4}{6}{6}{8}
  & \qldpc{100}{4}{4}{6}{3}
  & {\qldpc{74}{4}{4}{6}{3}}\\

\qldpc{65}{9}{4}{8}{3}
  & \qldpc{6844}{9}{5}{6}{8}
  & \qldpc{89}{9}{4}{6}{3}
  & {\qldpc{65}{9}{4}{6}{3}}\\

\qldpc{58}{16}{3}{8}{3}
  & \qldpc{5085}{16}{3}{6}{8}
  & \qldpc{136}{16}{4}{6}{3}
  & {\qldpc{58}{16}{3}{6}{3}}\\
\hline
\hline
\end{tabular}
\end{table}

\begin{table*}[ht]
\setlength{\tabcolsep}{20pt}  
\centering
\caption{Comparison of RL(6,3), RL(8,4), against SOTA on hypergraph product codes constructed from $n=30$ classical codes. Data is also shown in Fig.~\ref{fig:hastingsandn30}}
\label{tab:qldpc-comparison}
\begin{tabular}{l|aal}
\hline\hline
Base Code\ & RL(6,3) & RL(8,4) & SOTA \\
\hline
\qldpc{1741}{1}{30}{4}{2} & \qldpc{1741}{1}{30}{4}{2} & \qldpc{1741}{1}{30}{4}{2} & \qldpc{1741}{1}{30}{4}{2} \\
\qldpc{1684}{4}{20}{6}{19} & \qldpc{1802}{4}{20}{6}{3} & \qldpc{1741}{4}{20}{8}{4} & \qldpc{7444}{4}{36}{6}{3} \\
\qldpc{1629}{9}{16}{8}{15} & \qldpc{1745}{9}{16}{6}{3} & \qldpc{1745}{9}{16}{8}{4} & \qldpc{9389}{9}{29}{6}{3} \\
\qldpc{1576}{16}{16}{10}{15} & \qldpc{1930}{16}{16}{6}{3} & \qldpc{1690}{16}{16}{8}{4} & \qldpc{15496}{16}{33}{6}{3} \\
\qldpc{1525}{25}{15}{12}{15} & \qldpc{2125}{25}{15}{6}{3} & \qldpc{1997}{25}{15}{8}{4} & \qldpc{23557}{25}{33}{6}{3} \\
\qldpc{1476}{36}{14}{12}{15} & \qldpc{2330}{36}{14}{6}{3} & \qldpc{2066}{36}{14}{8}{4} & \qldpc{33300}{36}{36}{6}{3} \\
\qldpc{1429}{49}{12}{14}{11} & \qldpc{2137}{49}{12}{6}{3} & \qldpc{1765}{49}{12}{8}{4} & \qldpc{27637}{49}{28}{6}{3} \\
\qldpc{1384}{64}{12}{16}{15} & \qldpc{2624}{64}{12}{6}{3} & \qldpc{1954}{64}{12}{8}{4} & \qldpc{41504}{64}{33}{6}{3} \\
\qldpc{1341}{81}{12}{18}{15} & \qldpc{3161}{81}{12}{6}{3} & \qldpc{2153}{81}{12}{8}{4} & \qldpc{52853}{81}{33}{6}{3} \\
\qldpc{1300}{100}{11}{20}{14} & \qldpc{3412}{100}{11}{6}{3} & \qldpc{2228}{100}{11}{8}{4} & \qldpc{59908}{100}{35}{6}{3} \\
\qldpc{1261}{121}{10}{18}{13} & \qldpc{3673}{121}{10}{6}{3} & \qldpc{2173}{121}{10}{8}{4} & \qldpc{70373}{121}{32}{6}{3} \\
\qldpc{1224}{144}{9}{22}{13} & \qldpc{3944}{144}{9}{6}{3} & \qldpc{2120}{144}{9}{8}{4} & \qldpc{73800}{144}{28}{6}{3} \\
\qldpc{1189}{169}{8}{24}{11} & \qldpc{4225}{169}{8}{6}{3} & \qldpc{2069}{169}{8}{8}{4} & \qldpc{44785}{169}{22}{6}{3} \\
\qldpc{1156}{196}{8}{22}{11} & \qldpc{4706}{196}{8}{6}{3} & \qldpc{2146}{196}{8}{8}{4} & \qldpc{51940}{196}{23}{6}{3} \\
\qldpc{1125}{225}{8}{26}{11} & \qldpc{5417}{225}{8}{6}{3} & \qldpc{2493}{225}{8}{8}{4} & \qldpc{66346}{225}{23}{6}{3} \\
\qldpc{1096}{256}{7}{24}{11} & \qldpc{6626}{256}{7}{6}{3} & \qldpc{2440}{256}{7}{8}{4} & \qldpc{63496}{256}{22}{6}{3} \\
\qldpc{1069}{289}{6}{24}{7} & \qldpc{6529}{289}{6}{6}{3} & \qldpc{2389}{289}{6}{8}{4} & \qldpc{40757}{289}{15}{6}{3} \\
\qldpc{1044}{324}{6}{36}{9} & \qldpc{6660}{324}{7}{6}{3} & \qldpc{2612}{324}{6}{8}{4} & \qldpc{54612}{324}{18}{6}{3} \\
\qldpc{1021}{361}{6}{36}{9} & \qldpc{8761}{361}{6}{6}{3} & \qldpc{2845}{361}{6}{8}{4} & \qldpc{56293}{361}{20}{6}{3} \\
\qldpc{1000}{400}{5}{38}{8} & \qldpc{6928}{400}{5}{6}{3} & \qldpc{2792}{400}{5}{8}{4} & \qldpc{50128}{400}{20}{6}{3} \\
\qldpc{981}{441}{4}{32}{5} & \qldpc{7301}{441}{4}{6}{3} & \qldpc{2465}{441}{4}{8}{4} & \qldpc{10733}{441}{7}{6}{3} \\
\qldpc{964}{484}{4}{32}{5} & \qldpc{7684}{484}{4}{6}{3} & \qldpc{2692}{484}{4}{8}{4} & \qldpc{14692}{484}{7}{6}{3} \\
\qldpc{949}{529}{4}{28}{5} & \qldpc{9649}{529}{4}{6}{3} & \qldpc{2785}{529}{4}{8}{4} & \qldpc{12277}{529}{7}{6}{3} \\
\qldpc{936}{576}{4}{30}{5} & \qldpc{12456}{576}{4}{6}{3} & \qldpc{3026}{576}{4}{8}{4} & \qldpc{17960}{576}{11}{6}{3} \\
\qldpc{925}{625}{3}{32}{5} & \qldpc{6925}{625}{3}{6}{3} & \qldpc{3277}{625}{3}{8}{4} & \qldpc{15625}{625}{9}{6}{3} \\
\qldpc{916}{676}{2}{54}{1} & \qldpc{1396}{676}{2}{6}{3} & \qldpc{1396}{676}{2}{8}{4} & \qldpc{2294}{676}{2}{6}{2} \\
\qldpc{909}{729}{2}{56}{1} & \qldpc{1469}{729}{2}{6}{3} & \qldpc{1469}{729}{2}{8}{4} & \qldpc{3809}{729}{2}{6}{2} \\
\qldpc{904}{784}{2}{58}{1} & \qldpc{1544}{784}{2}{6}{3} & \qldpc{1544}{784}{2}{8}{4} & \qldpc{3920}{784}{2}{6}{2} \\
\qldpc{901}{841}{2}{60}{1} & \qldpc{1621}{841}{2}{6}{3} & \qldpc{1621}{841}{2}{8}{4} & \qldpc{4033}{841}{2}{6}{2} \\
\qldpc{900}{900}{1}{2}{1} & \qldpc{900}{900}{1}{2}{1} & \qldpc{900}{900}{1}{2}{1} & \qldpc{900}{900}{1}{2}{1} \\
\hline\hline
\end{tabular}
\end{table*}

\begin{table}[ht]
\centering
\caption{Some examples of comparisons between a RL produced code and various codes produced by the method from Sabo et al.~\cite{sabo_weight-reduced_2024} with different base codes. On each code the highest distance over 100 tries was taken for the method by Ref.~\cite{sabo_weight-reduced_2024}.}
\label{tab:rl-sabo-comparison}
\begin{tabular}{lll}
\hline
\hline
\textbf{} & Reduced Code & Base Code \\
\hline
RL(6,3) &
\qldpc{2257}{25}{15}{6}{3} &
\qldpc{1525}{25}{15}{12}{15} \\
\hline
\multirow{6}{*}{Sabo et al.~\cite{sabo_weight-reduced_2024}} &
\begin{tabular}[t]{@{}l@{}}
\qldpc{2257}{25}{13}{6}{3}\\
\qldpc{3457}{25}{15}{6}{3}\\
\qldpc{4337}{25}{17}{6}{3}\\
\qldpc{4153}{25}{16}{6}{3}\\
\qldpc{4337}{25}{16}{6}{3}\\
\qldpc{4525}{25}{17}{6}{3}
\end{tabular}
&
\begin{tabular}[t]{@{}l@{}}
\qldpc{277}{25}{6}{6}{10}\\
\qldpc{325}{25}{7}{12}{7}\\
\qldpc{377}{25}{9}{12}{7}\\
\qldpc{433}{25}{8}{10}{7}\\
\qldpc{493}{25}{8}{10}{7}\\
\qldpc{557}{25}{8}{10}{7}
\end{tabular}\\
\hline
\hline
\end{tabular}
\end{table}
   
\begin{table}[H]
\centering
\caption{Qubit overheads when allowing small reductions in distance at $w=6$, $q=3$. Data is also shown in Fig.~\ref{fig:reducedistance}}
\label{tab:qldpc_overheads}
\begin{tabular}{cccccc}
\hline
\hline
Base Code & \textbf{$d$} & \textbf{$d-1$} & \textbf{$d-2$} & \textbf{$d-3$} \\
\hline
\qldpc{1525}{25}{15}{12}{15} & 600 & 293 & 112 & 0 \\
\qldpc{1476}{36}{14}{12}{15} & 854 & 590 & 342 & 110 \\
\qldpc{1429}{49}{12}{14}{11} & 708 & 456 & 108 & 0 \\
\qldpc{1282}{64}{12}{16}{15} & 1240 & 1098 & 826 & 570 \\
\qldpc{1241}{81}{11}{18}{14} & 1764 & 1320 & 912 & 660 \\
\qldpc{1108}{100}{10}{16}{13} & 2304 & 1392 & 1120 & 864 \\
\hline
\hline
\end{tabular}
\end{table}

\subsection{Figures}
\begin{figure*}[htbp]
    \centering
    \subfloat[\label{fig:spectralgap}]{
        \includegraphics[width=0.45\linewidth]{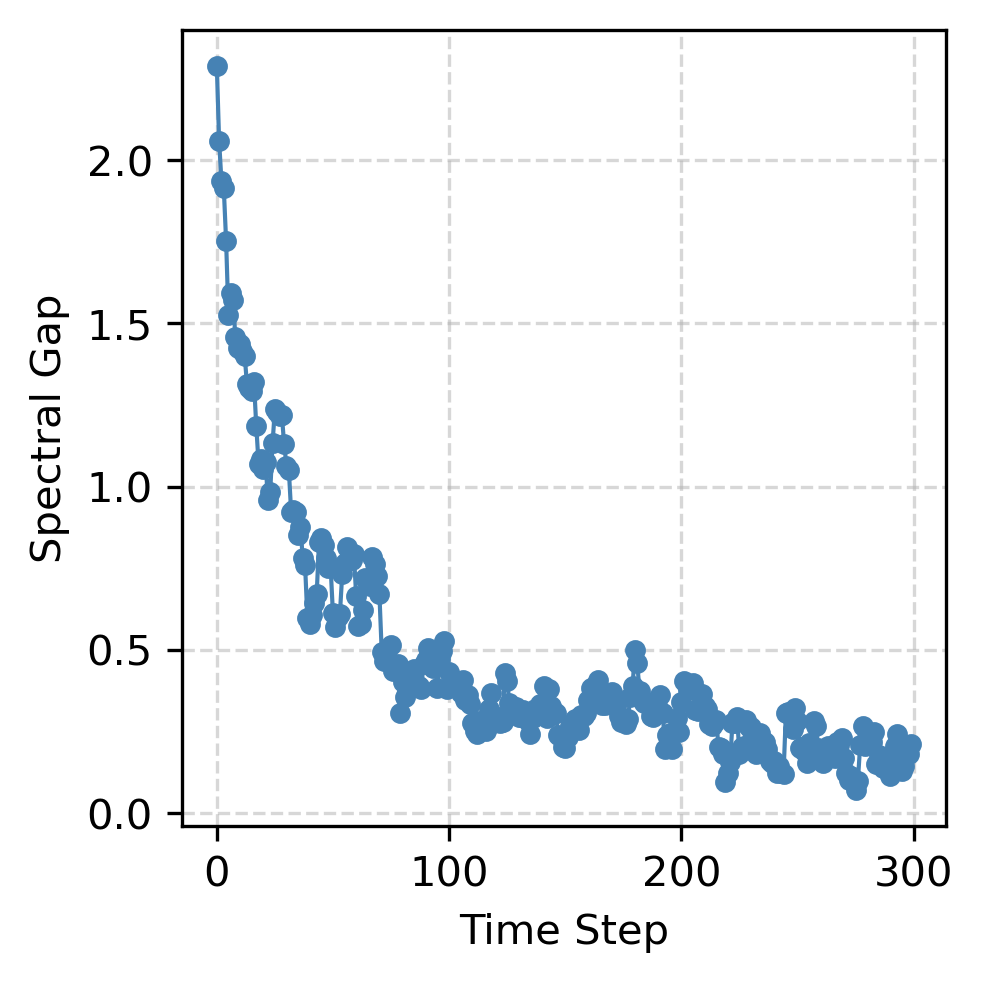}
    }
    \quad
    \subfloat[\label{fig:eigenvaluespectrogram}]{
        \includegraphics[width=0.45\linewidth]{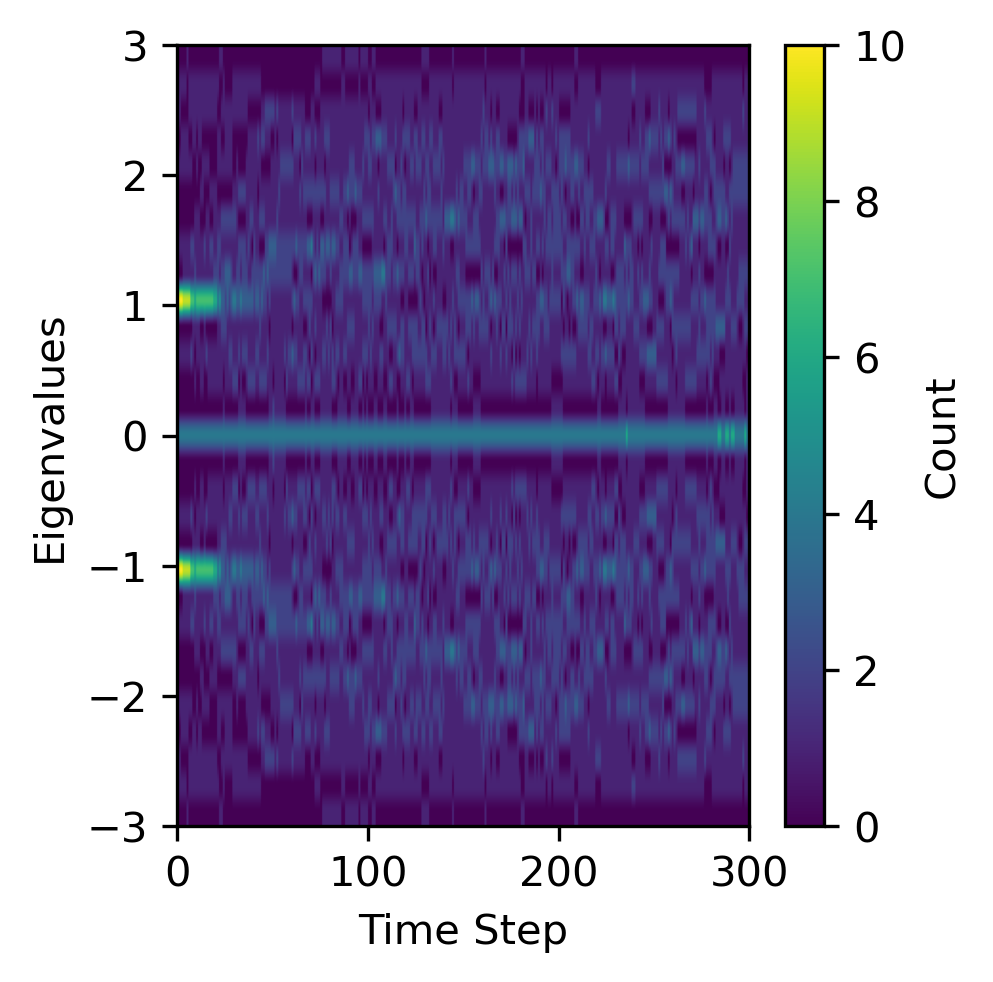}
    }
    \caption{%
    (a)~Spectral gap evolution.
    Our RL agent causes the Tanner graphs to rapidly lose their expander properties, with the gap \(\lambda_1 - \lambda_2\) stabilizing around \(\le 0.5\). 
    This happens simultaneously to reduction of weight and degree in the Tanner graph.
    (b)~Tanner graph eigenvalues evolution.
    The eigenvalues begin concentrated around \(+1\) and \(-1\), and spread out quickly. 
    It is interesting that the codes our agent finds are significantly less structured and tend to have nearly random eigenvalue distributions. 
    This suggests that our agent finds codes largely outside the realm of theoretical constructions, 
    which often tend to rely on expansion-related arguments, although this is also at the cost of worse performance on message-passing-based decoding.%
    }
    \label{fig:merged_spectra}
\end{figure*}

\begin{figure*}
    \centering
    \includegraphics[width=1\linewidth]{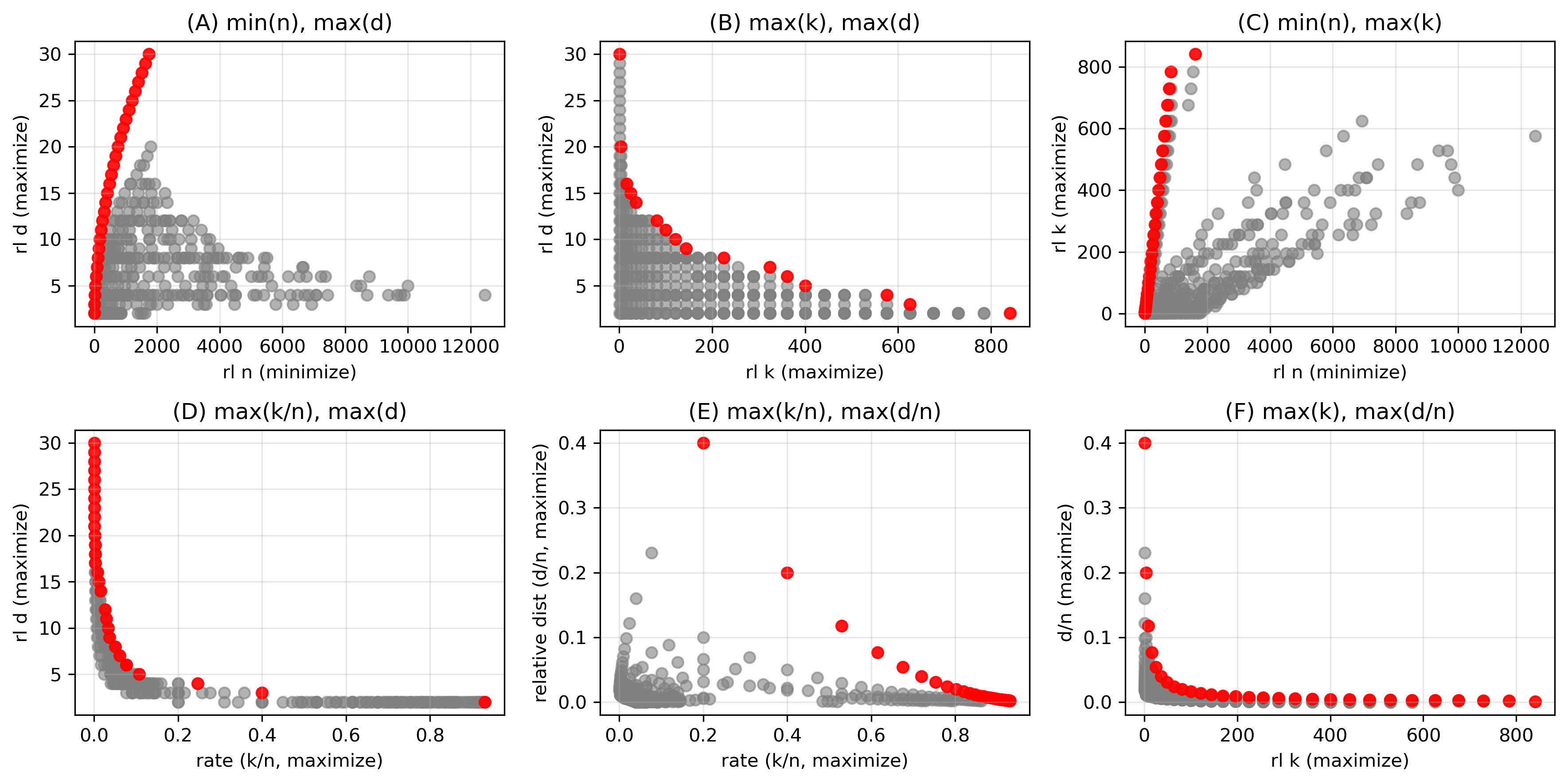}
    \caption{Pareto fronts of $n,k,d$ parameters.
    The plotted Pareto fronts show locally optimal \((n,k,d)\) codes found by our RL agent. We observe there is considerable opportunity to improve \(k/n\) and \(d/n\) in the weight-reduction setting, especially since hypergraph product code codes cannot reach certain theoretical bounds. 
    \textit{Note:} The 10 additional points discussed in the main text are omitted from these plots.}
    \label{fig:paretofronts}
\end{figure*}

\begin{figure}[htbp]
    \centering
    \subfloat[]{
        \includegraphics[width=0.45\linewidth]{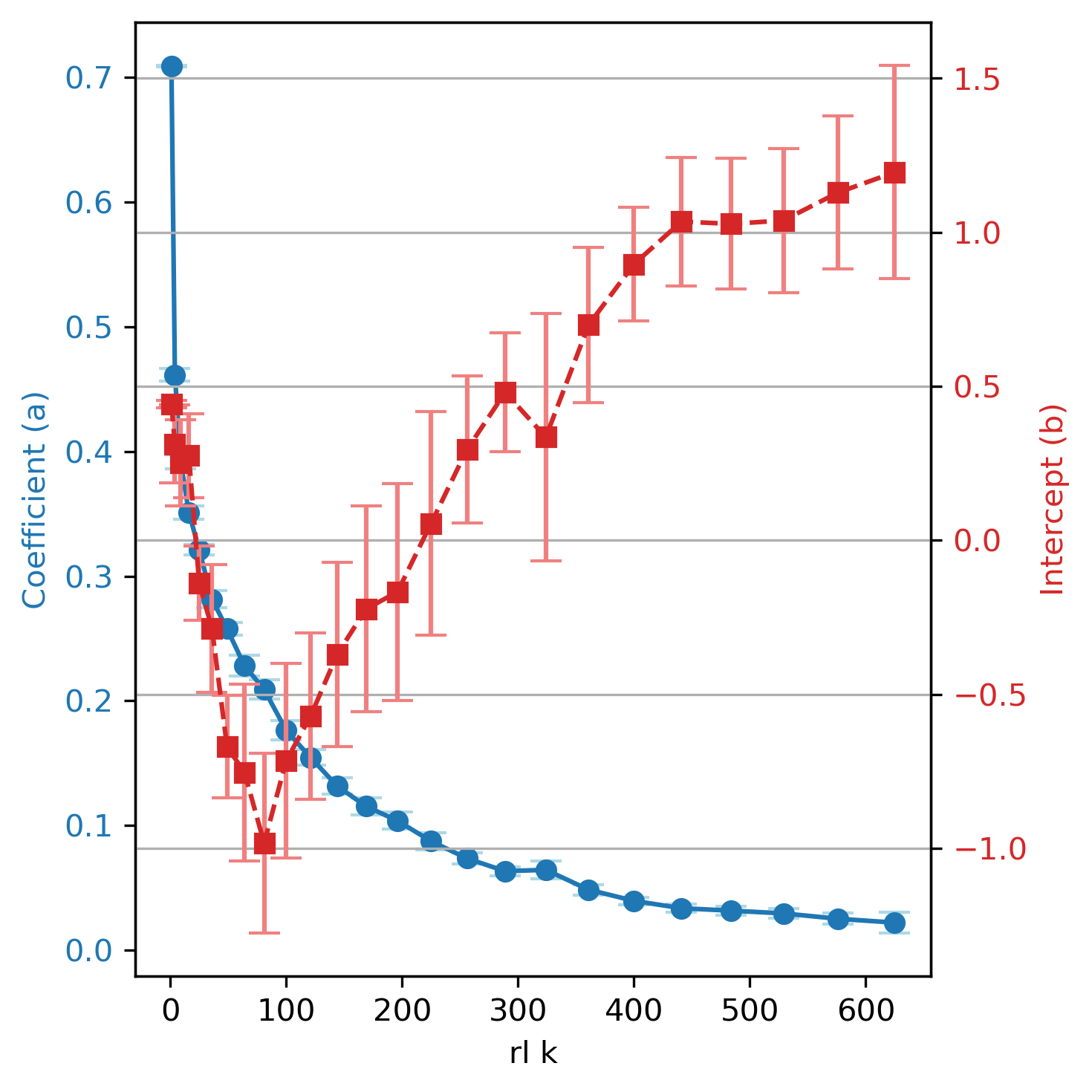}%
        \label{fig:metaanalysisd}
    }
    \quad
    \subfloat[]{
        \includegraphics[width=0.45\linewidth]{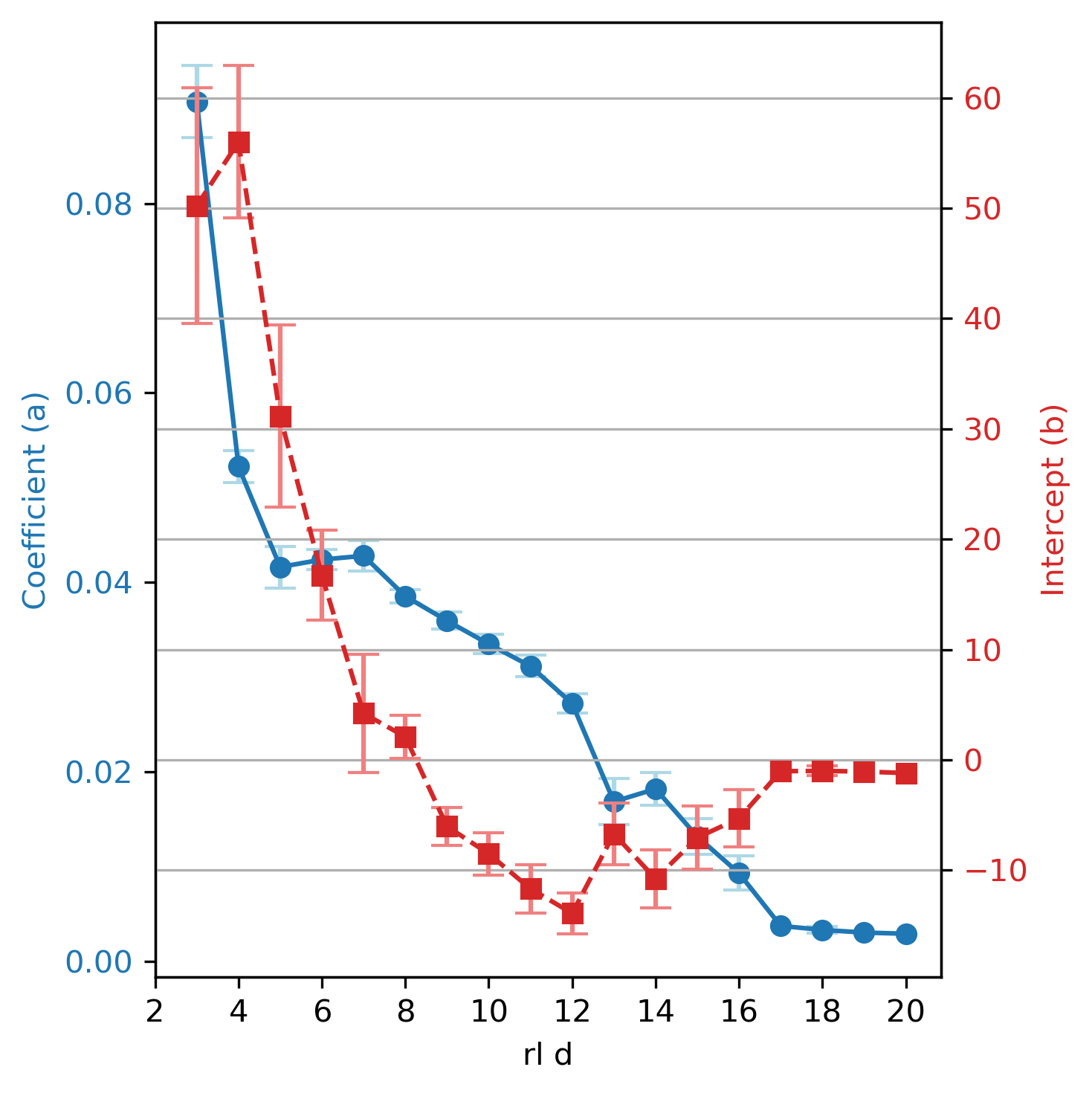}%
        \label{fig:metaanalysisk}
    }
    \\[1em]
    \subfloat[]{
        \includegraphics[width=0.45\linewidth]{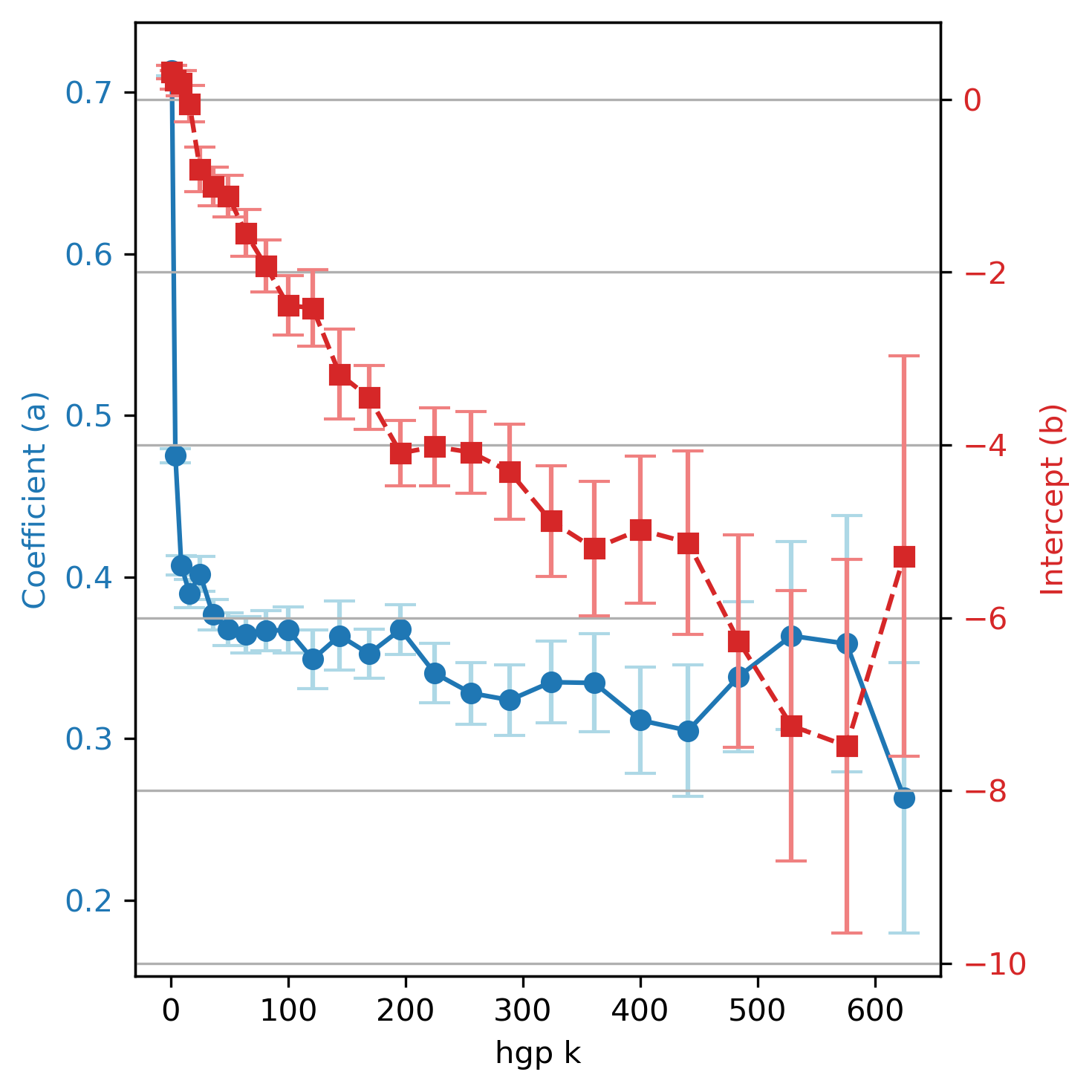}%
        \label{fig:metaanalysishgpd}
    }
    \quad
    \subfloat[]{
        \includegraphics[width=0.45\linewidth]{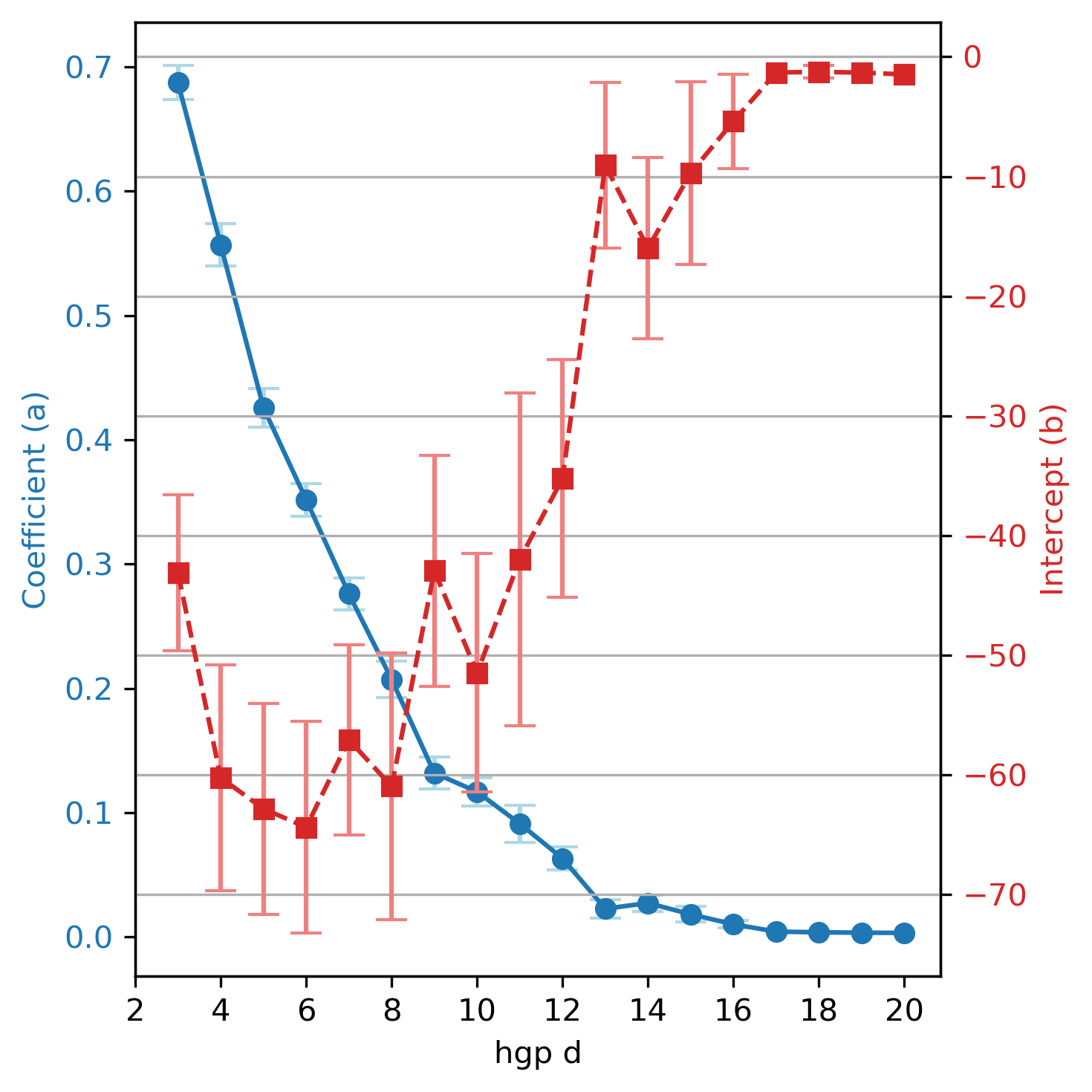}%
        \label{fig:metaanalysishgpk}
    }
    \caption{%
    (a)~Meta-analysis of square-root regressions on RL $n$ vs.\ $d$ at various $k$ values.
    Each curve shows how the best-fit $n$ vs. \(d\) relationship changes at unique \(k\). 
    After weight reduction, the scaling factor of \(d\) to \(n\) decreases with \(k\), eventually flattening at high \(k\). 
    In contrast, hypergraph product codes prior to weight reduction maintain a roughly constant scaling factor for all \(k\). (b)~Meta-analysis of linear regressions on RL $n$ vs.\ $k$ at various $d$ values.
    Each line shows how $n$ scales with $k$ for fixed $d$. 
    The coefficient tends to decrease both before and after weight reduction (c)~Meta-analysis of square-root regressions for hypergraph product codes on $n$ vs.\ $d$ for varying $k$.
    Before weight reduction, the hypergraph product codes do not maintain low $w,q$; hence their scaling factors remain fairly constant across different $k$ values.
    These points are also optimal for hypergraph product codes in terms of $n,k,d$ but still exhibit growth in weight and degree not seen in RL-generated codes. 
    (d)~Meta-analysis of linear regressions for hypergraph product codes on $n$ vs.\ $k$ for varying $d$. Coefficients show a decreasing trend, although beginning at larger values than (b).
    }
    \label{fig:metaanalysis2x2}
\end{figure}

\begin{figure}
    \centering
    \includegraphics[width=0.92\linewidth]{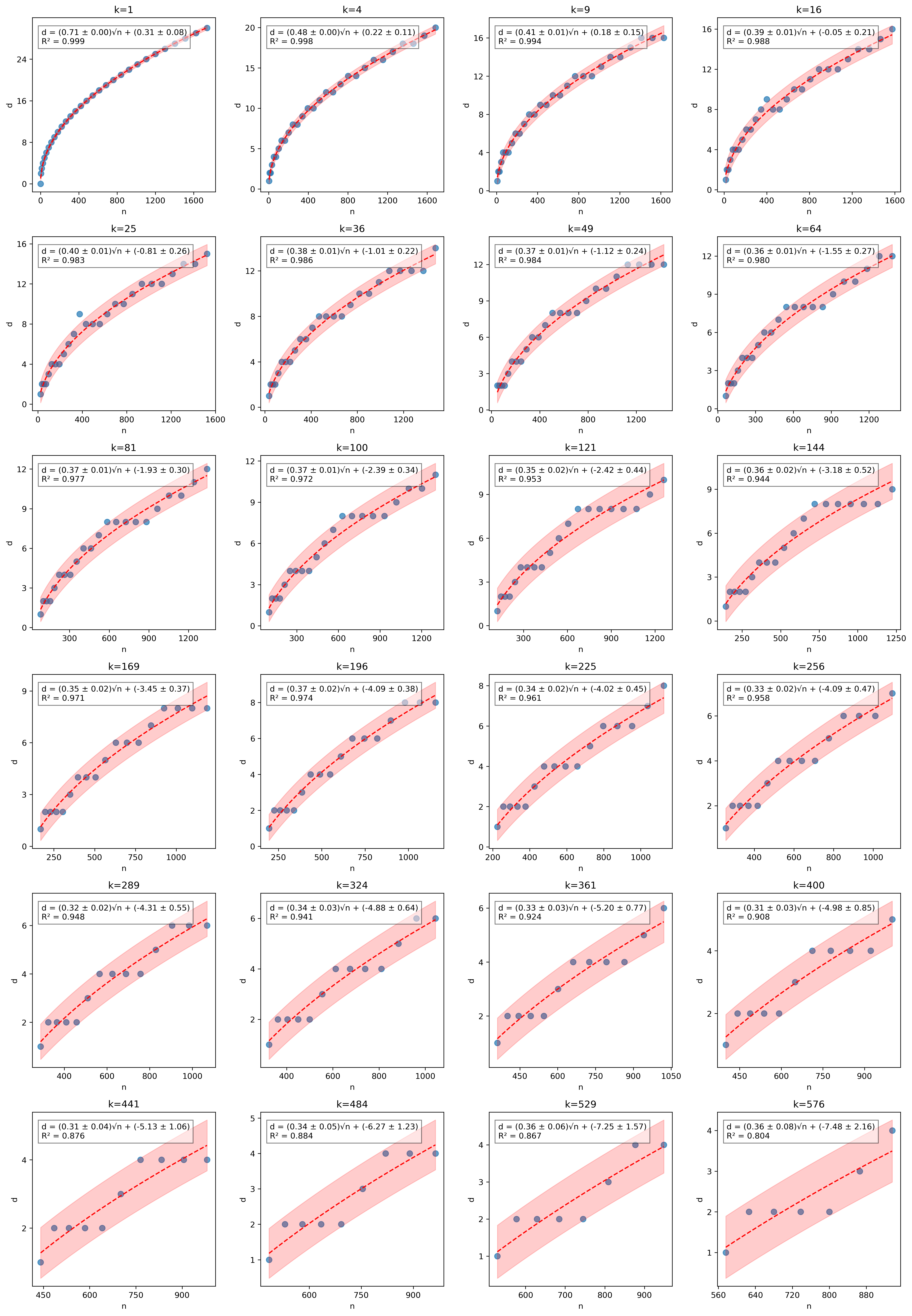}
    \caption{Regressions of $n$ and $d$ for hypergraph product codes (no weight reduction). The coefficient in the $n$ vs. \(d\) regressions remains mostly constant as \(k\) varies, and there is an observable auto-correlation in the placements of these points. At large \(k\), both the range of \(d\) and the associated \(r^2\) values tend to drop, reflecting increased uncertainty at high code rates.}
    \label{fig:hgpregressions}
\end{figure}

\begin{figure}
    \centering
    \includegraphics[width=0.92\linewidth]{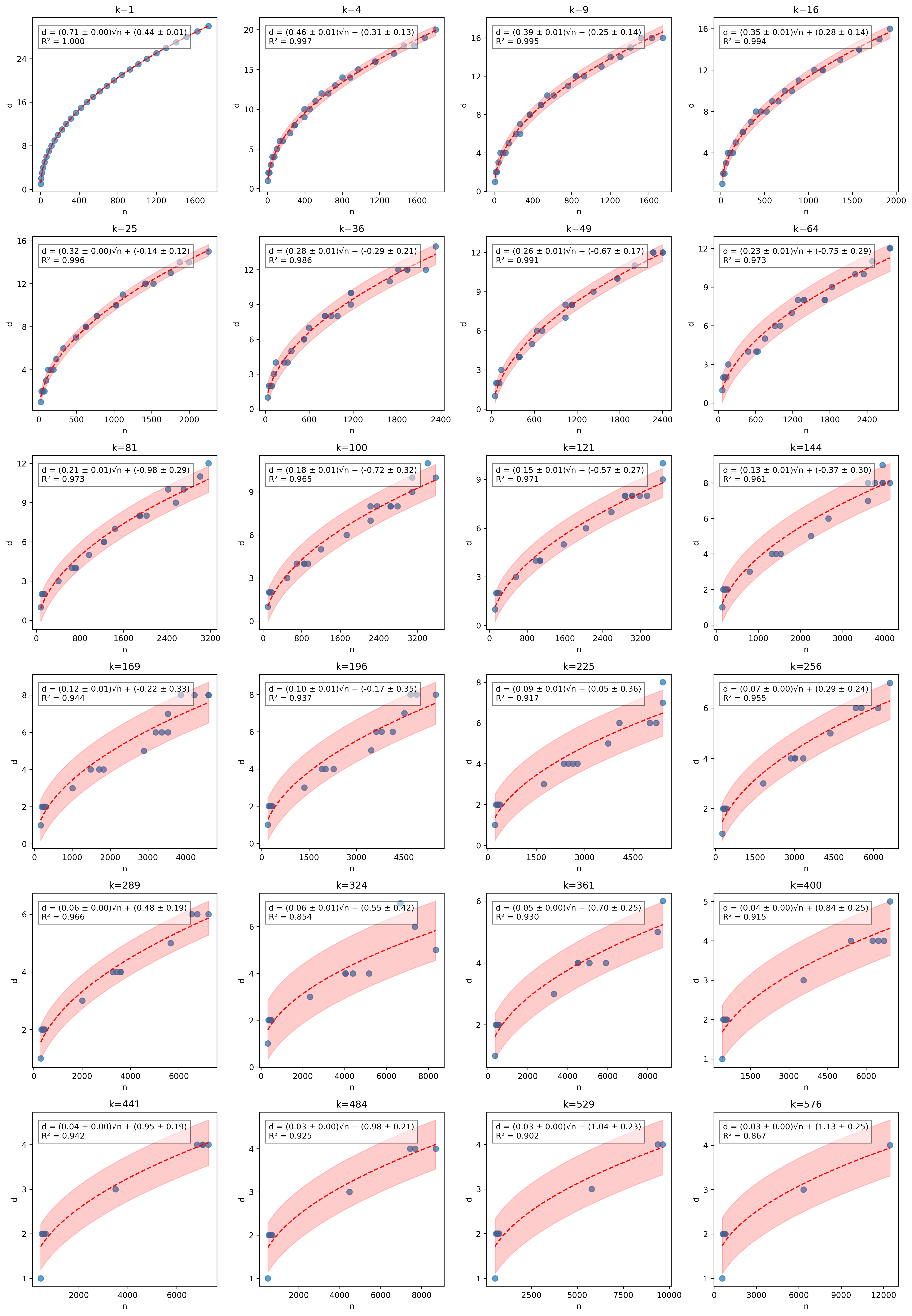}
    \caption{Regressions of $n$ and $d$ for RL-generated codes (after weight reduction). Once weight and degree are reduced, the coefficients typically decrease. 
    At large \(k\), both the range of \(d\) and the associated \(r^2\) values tend to drop, reflecting increased uncertainty at high code rates.
    Also, some inefficiencies in the RL optimization process can improve the \(r^2\) fit for certain \(k\) values.
    }
    \label{fig:rlregressions}
\end{figure}

\begin{figure}
    \centering
    \includegraphics[width=1\linewidth]{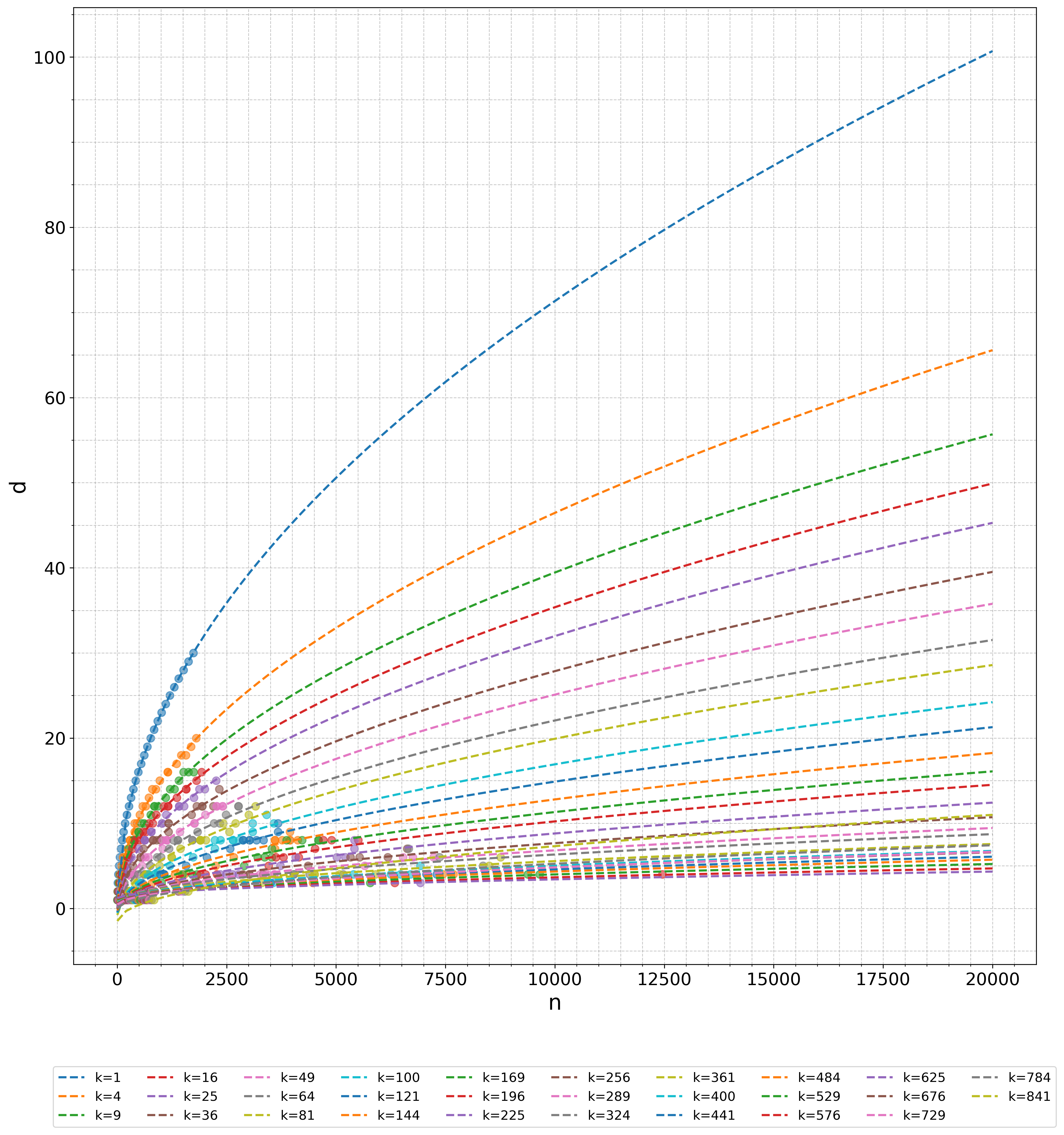}
    \caption{Extrapolations of $n$ vs. $d$ regressions to $n=20,000$ for RL-generated (6,3) codes from the HGP-30 regime. Error bars shown in Fig.~\ref{fig:rlregressions} are omitted for ease of visualization. 
    }
    \label{fig:extrapolations}
\end{figure}

Spectral properties of the codes produced by our RL agent are demonstrated in Fig.~\ref{fig:spectralgap} and Fig.~\ref{fig:eigenvaluespectrogram}. Pareto fronts of our (6,3) codes in terms of $n,k,d,k/n,d/n$ are shown in Fig.~\ref{fig:paretofronts}. Further analysis of weight reduction trade-offs are made through a meta analysis of $n,k,d$ in Fig.~\ref{fig:metaanalysis2x2}, with individual regressions for $n$ vs. $d$ at different k values shown in Fig.~\ref{fig:hgpregressions}, Fig.~\ref{fig:rlregressions}, as well as an extrapolation for possible code parameters obtainable by our RL framework up to $n=20,000$ in Fig.~\ref{fig:extrapolations}.

\end{document}